\documentclass[letterpaper, 11 pt,onecolumn,draftclsnofoot]{IEEEtran}
\usepackage[T1]{fontenc}
\usepackage[latin1]{inputenc}
\usepackage{float}
\usepackage{amsmath}
\usepackage[noadjust]{cite}
\usepackage{amssymb}
\usepackage{enumerate}
\usepackage{amsthm}
\usepackage{graphics,epsfig}
\usepackage{subfigure,epsf,pstricks}
\usepackage{bm}
\usepackage{url}

\floatstyle{ruled}
\newfloat{algorithm}{tbp}{loa}
\floatname{algorithm}{Algorithm}

\usepackage{algorithmic}
\algsetup{linenodelimiter= }

\def\bfe{{\mathbf e}}

\def\bfv{{\mathbf v}}

\def\bfx{{\mathbf x}}
\def\bfy{{\mathbf y}}
\def\bfz{{\mathbf z}}

\def\bfX{{\mathbf X}}

\title{On the Performance of Sparse Recovery via $\ell_p$-minimization ($0 \leq p \leq 1$)}

\author{Meng Wang \ \ \ \ \ Weiyu Xu \ \ \ \ \ Ao Tang
\\
School of ECE, Cornell University, Ithaca, NY 14853, USA }

\usepackage{graphicx}
\usepackage{epstopdf}
\newtheorem{theorem}{Theorem}
\newtheorem{lemma}{Lemma}
\newtheorem{cor}{Corollary}

\newtheorem{prop}{Proposition}

\begin{document}

\maketitle \thispagestyle{empty} \pagestyle{empty}

\begin{abstract}
It is known that a high-dimensional sparse vector $\bfx^*$ in
$\mathcal{R}^n$ can be recovered from low-dimensional measurements
$\bfy=A\bfx^*$ where $A^{m \times n}$($m<n$) is the measurement matrix.
In this paper, we investigate 
the recovering ability of $\ell_p$-minimization ($0\leq p \leq 1$)
as $p$ varies, where $\ell_p$-minimization returns a vector with the
least $\ell_p$ ``norm'' among all the vectors $\bfx$ satisfying
$A\bfx=\bfy$. Besides analyzing the performance of strong recovery
where $\ell_p$-minimization is required to recover all the sparse
vectors up to certain sparsity, we also for the first time analyze
the performance of ``weak'' recovery of $\ell_p$-minimization
($0\leq p<1$) where the aim is to recover all the sparse vectors on
one support with fixed sign pattern. When $\alpha (:=\frac{m}{n})
\rightarrow 1$, we provide sharp thresholds of the sparsity ratio
that differentiates the success and failure via
$\ell_p$-minimization. For strong recovery, the threshold strictly
decreases from 0.5 to 0.239 as $p$ increases from 0 to 1.
Surprisingly, for weak recovery, the threshold is $2/3$ for all $p$
in $[0,1)$, while the threshold is 1 for $\ell_1$-minimization. We
also explicitly demonstrate that $\ell_p$-minimization ($p<1$) can
return a denser solution than $\ell_1$-minimization. For any
$\alpha<1$, we provide bounds of sparsity ratio for strong recovery
and weak recovery respectively below which $\ell_p$-minimization
succeeds with overwhelming probability. Our bound of strong recovery
improves on the existing bounds when $\alpha$ is large. In
particular, regarding the recovery threshold, this paper argues that
$\ell_p$-minimization has a higher threshold with smaller $p$ for
strong recovery; the threshold is the same for all $p$ for sectional
recovery; and $\ell_1$-minimization can outperform
$\ell_p$-minimization for weak recovery. These are in contrast to
traditional wisdom that $\ell_p$-minimization, though
computationally more expensive, always has better sparse recovery
ability than $\ell_1$-minimization since it is closer to
$\ell_0$-minimization. Finally, we provide an intuitive explanation
to our findings. Numerical examples are also used to unambiguously
confirm and illustrate the theoretical predictions.
\end{abstract}

\section{Introduction} \label{sec:intro}
We consider recovering a vector $\bfx$ in $\mathcal{R}^n$ from an
$m$-dimensional measurement $\bfy=A\bfx$, where $A^{m \times n}$($m
< n$) is the measurement matrix. Obviously, given $\bfy$ and $A$,
$A\bfx=\bfy$ is an underdetermined linear system and admits an
infinite number of solutions. However, if $\bfx$ is sparse, i.e. it
only has a small number of nonzero entries compared with its
dimension, one can actually recover $\bfx$ from $\bfy$. This topic
is known as \textit{compressed sensing} and draws much attention
recently, for example,
\cite{CaT05}\cite{CaT06}\cite{DoT05}\cite{Don06}.

Given $\bfx \in \mathcal{R}^n$, its support $T$ is defined as
$T=\{i\in\{1,...,n\}: x_i \neq 0 \}$. The cardinality $|T|$ of set
$T$ is the sparsity of $\bfx$, which also equals to the $\ell_0$
norm $\|\bfx\|_0:=|\{i:x_i \neq 0\}|$. We say $\bfx$ is $\rho
n$-sparse
if $|T| = \rho n$ for some $\rho<1$. 
Given the measurement $\bfy$ and the measurement matrix $A$,
together with the assumption that $\bfx$ is sparse, one natural
estimate of $\bfx$ is the vector with the least $\ell_0$ norm that
can produce the measurement $\bfy$. Mathematically, to recover
$\bfx$, we solve the following $\ell_0$-minimization problem:
\begin{equation}\label{eqn:l0}
\min \limits_{\bfx \in \mathcal{R}^n} \|\bfx \|_0 \quad
\textrm{s.t.} \quad A\bfx=\bfy.
\end{equation}
However, (\ref{eqn:l0}) is combinatorial and computationally
intractable, and one commonly used approach is to solve a closely
related $\ell_1$-minimization problem:
\begin{equation}\label{eqn:l1}
\min \limits_{\bfx \in \mathcal{R}^n} \|\bfx \|_1 \quad
\textrm{s.t.} \quad A\bfx=\bfy,
\end{equation}
where $\|\bfx\|_1:= \sum_i |x_i|$. (\ref{eqn:l1}) is a convex
problem and can be recast as a linear program, thus can be solved
efficiently. Conditions under which (\ref{eqn:l1}) can successfully
recover $\bfx$ have been extensively studied in the literature of
compressed sensing. For example, one widely known sufficient
condition is the Restricted Isometry Property (RIP)
\cite{CRT06}\cite{CaT05}\cite{CaT06}.

Among the explosion of research on compressed sensing
(\cite{BDDW08}\cite{BeI08}\cite{BEZ08}\cite{CDD09}\cite{HN06}\cite{WM08}\cite{XuH07}),
recently, there has been great research interest in recovering
$\bfx$ by $\ell_p$-minimization for $0<p<1$
(\cite{Chartrand07}\cite{Chartrand072}\cite{CY08}\cite{DG09}\cite{FL09}\cite{SCY08}\cite{BGIKS08})
as follows,
\begin{equation}\label{eqn:lp}
\min \limits_{\bfx \in \mathcal{R}^n} \|\bfx \|_p \quad
\textrm{s.t.} \quad A\bfx=\bfy.
\end{equation}
Recall that $ \|\bfx\|_p^p:=\left(\sum_i |x_i|^p\right)$ for $p>0$.
Though $\|\cdot\|_p$ does not actually define a \textit{norm} as it
violates the triangular inequality, $\|\cdot\|_p^p$ follows the
triangular inequality. We say $\bfx$ can be recovered by
$\ell_p$-minimization if and only if it
is the unique solution to (\ref{eqn:lp}). 
%
%
%
(\ref{eqn:lp}) is non-convex, and thus it is generally hard to
compute the global minimum.
\cite{Chartrand07}\cite{Chartrand072}\cite{CY08} employ heuristic
algorithms to compute a local minimum of (\ref{eqn:lp}) and show
numerically that these heuristics can indeed recover sparse vectors,
and the support size of these vectors can be larger than that of the
vectors recoverable from $\ell_1$-minimization.
Then the question is what is the relationship between the sparsity
of a vector and the successful recovery with $\ell_p$-minimization
($p<1$)? How sparse should a vector be so that $\ell_p$-minimization
can recover it? \cite{GN03} shows the sparsity up to which
$\ell_p$-minimization can successfully recover all the sparse
vectors at least does not decrease as $p$ decreases. \cite{SCY08}
provides a sufficient condition for successful recovery via
$\ell_p$-minimization based on Restricted Isometry Constants and
provides a lower bound of the support size up to
which $\ell_p$-minimization can recover all such sparse vectors. 
\cite{FL09} improves this bound by considering a generalized version
of RIP condition, and \cite{BCT09} numerically calculates this bound.  


Here are the main contributions of this paper. For strong recovery
where $\ell_p$-minimization needs to recover all the vectors up to a
certain sparsity, we provide a sharp threshold $\rho^*(p)$ of the
ratio of the support size to the dimension which differentiates the
success and the failure of $\ell_p$-minimization when $\alpha (=
\frac{m}{n})\rightarrow 1$. This is an exact threshold compared with
a lower bound of successful recovery in previous results. When
$\rho$ increases from 0 to 1, $\rho^*(p)$ decreases from 0.5 to
0.239. This coincides with the intuition that the performance of
$\ell_p$-minimization is improved when $p$ decreases. When $\alpha
<1$ is fixed, we provide a positive bound $\rho^*(\alpha,p)$ for all
$\alpha \in (0,1)$ and all $p\in (0,1]$ of strong recovery such that
with a Gaussian measurement matrix $A^{m \times n}$,
$\ell_p$-minimization can recover all the $\rho^*(\alpha,p)n$-sparse
vectors with overwhelming probability. $\rho^*(\alpha,p)$ improves
on the existing bound in large $\alpha$ region.

We also analyze the performance of $\ell_p$-minimization for
\textit{weak} recovery where we need to recover all the sparse
vectors on one support with one sign pattern. To the best of our
knowledge, there is no existing result in this regard for $p<1$. We
characterize the successful weak recovery through a necessary and
sufficient condition regarding the null space of the measurement
matrix. When $\alpha \rightarrow 1$, we provide a sharp threshold
$\rho_w^*(p)$ of the ratio of the support size to the dimension
which differentiates the success and the failure of
$\ell_p$-minimization. The weak threshold indicates that if we would
like to recover every vector over one support with size less than
$\rho_w^*(p) n$ and with one sign pattern, (though the support and
sign patterns are not known a priori), and we generate a random
Gaussian measurement matrix independently of the vectors, then with
overwhelmingly high probability, $\ell_p$-minimization will recover
all such vectors regardless of the amplitudes of the entries of a
vector. For $\ell_1$-minimization, given a vector, if we randomly
generate a Gaussian matrix and apply $\ell_1$-minimization, then its
recovering ability observed in simulation exactly captures the weak
recovery threshold, see \cite{Donoho06}\cite{DoT05}.
Interestingly, we prove that the weak threshold $\rho^*_w(p)$ is 2/3
for all $p \in [0,1)$, and is lower than the weak threshold of
$\ell_1$-minimization, which is 1. Therefore, $\ell_1$-minimization
outperforms $\ell_p$-minimization for all $p\in[0,1)$ if we only
need to recover
sparse vectors on one support with one sign pattern. 
 We also explicitly show that $\ell_p$-minimization ($p \in(0,1)$) can return a vector
denser than the original sparse vector while $\ell_1$-minimization
successfully recovers the sparse vector. Finally, for every $\alpha <1$, we
provide a positive bound $\rho^*_w(\alpha,p)$ such that
$\ell_p$-minimization successfully recovers all the
$\rho^*_w(\alpha,p)n$-sparse vectors on one support with one sign
pattern.

The rest of the paper is organized as follows. We introduce the null
space condition of successful $\ell_p$-minimization in Section
\ref{sec:null}. We especially define the successful weak recovery
for $p<1$ and provide a necessary and sufficient condition. We use
an example to illustrate that the solution of $\ell_1$-minimization
can be sparser than that of $\ell_p$-minimization ($p \in(0,1)$).
Section \ref{sec:limit} provides thresholds of the sparsity ratio of
the successful recovery via $\ell_p$-minimization for all $p
\in[0,1]$ both in strong recovery and in weak recovery when the
measurement matrix is random Gaussian matrix and $\alpha \rightarrow
1$. For $\alpha <1$, Section \ref{sec:finite} provides bounds of
sparsity ratio below which $\ell_p$-minimization is successful in
the strong sense and in the weak sense respectively. We compare the
performance of $\ell_p$-minimization ($p<1$) and the performance of
$\ell_1$-minimization in Section \ref{sec:lpl1} and provide
numerical results in Section \ref{sec:simu}. Section
\ref{sec:conclusion} concludes the paper.

\section{Successful Recovery of $\ell_p$-minimization}\label{sec:null}
We first introduce the null space characterization of the
measurement matrix $A$ to capture the successful recovery via
$\ell_p$-minimization ($p\in[0,1]$). Besides the strong recovery
that has been studied in
\cite{BCT09}\cite{CDD09}\cite{FL09}\cite{Fuchs04}\cite{GN03}\cite{SCY08}\cite{SXH08},
we especially provide a necessary and sufficient condition for the
success of \textit{weak} recovery in the sense that
$\ell_p$-minimization only needs to recover all the sparse vectors
on one support with one sign pattern. For example, in practice,
given an unknown vector to recover, we randomly generate a
measurement matrix and solve the $\ell_1$-minimization problem, the
simulation result of recovery performance with respect to the
sparsity of the vector indeed represents the performance of weak
recovery.

Given a measurement matrix $A^{m \times n}$, let $B^{n \times
(n-m)}$ denote a basis of the null space of $A$, then we have
$AB=\bf{0}$. Let $B_i$ ($i\in\{1,...,n\}$) denote the
$i^{\textrm{th}}$ row of $B$. Let $B_T$ denote the submatrix of $B$
with $T \subseteq \{1,...,n\}$ as the set of row indices. In this
paper, we will study the sparse recovery property of
$\ell_p$-minimization by analyzing the null space of $A$.

We first state  the null space condition for the success of strong
recovery via $\ell_p$-minimization (\cite{EB02}\cite{GN03}) in the
sense that $\ell_p$-minimization should recover all the sparse
vectors up to a certain sparsity.
\begin{theorem}[\cite{EB02}\cite{GN03}]\label{thm:slp}
$\bfx$ is the unique solution to $\ell_p$-minimization problem $(0
\leq p \leq 1)$ for every vector $\bfx$ up to $\rho n$-sparse if and
only if
\begin{equation}\label{eqn:slp}
\|B_T\bfz\|_p^p <  \|B_{T^c}\bfz\|_p^p
\end{equation}
for every non-zero $\bfz \in \mathcal{R}^{n-m}$, and every support
$T$ with $|T| \leq \rho n$.
\end{theorem}

One important property is that if the condition (\ref{eqn:slp}) is
satisfied for some $0<p \leq 1$, then it is also satisfied for all
$q \in [0,p]$ (\cite{DG09}\cite{GN07}). Therefore, if
$\ell_p$-minimization could recover all the $\rho n$-sparse vectors
$\bfx$, then $\ell_q$-minimization ($0 \leq q \leq p$) could also
recover all the $\rho n$-sparse vectors. Intuitively, the strong
recovery performance of $\ell_q$-minimization should be at least as
good as that of $\ell_p$-minimization when $0\leq q<p\leq 1$.

\subsection{Weak recovery for $\ell_p$-minimization}

Though $\ell_p$-minimization ($p<1$) should be at least as good as
$\ell_1$-minimization for strong recovery, the argument may not be
true for weak recovery.

We first state the null space condition for successful weak recovery
via $\ell_1$-minimization as follows, (see
\cite{DH01}\cite{GN03}\cite{Stojnic09}\cite{XH08}\cite{Zhang06} for
this result.)
\begin{theorem}\label{thm:wl1}
For every $\bfx \in \mathcal{R}^n$ on some support $T$ with the same
sign pattern, $\bfx$ is always the unique solution to
$\ell_1$-minimization problem (\ref{eqn:l1}) if and only if
\begin{equation}\label{eqn:wl1}
\|B_{T^-}\bfz\|_1 < \|B_{T^c}\bfz\|_1+\|B_{T^+}\bfz\|_1
\end{equation}
holds for all non-zero $\bfz \in \mathcal {R}^{n-m}$ where $T^-=\{ i
\in T: B_i \bfz x_i<0 \}$, and $T^+=\{ i \in T: B_i \bfz x_i\geq 0
\}$
\end{theorem}

Note that for every vector $\bfx$ on a fixed support $T$ with a
fixed sign pattern, the condition to successfully recover it via
$\ell_1$-minimization is the same, as stated in Theorem
\ref{thm:wl1}. However, the condition of successful recovery via
$\ell_p$-minimization ($0\leq p<1$) varies for different sparse
vectors even if they have the same support and the same sign
pattern. In other words, the recovery condition depends on the
amplitudes of the entries of the vector. Here we consider the worst
case scenario for weak recovery in the sense that the recovery via
$\ell_p$-minimization is defined to be ``successful'' if it can
recover \textit{all} the vectors on a fixed support with a fixed
sign pattern. The null space condition for weak recovery in this
definition via $\ell_1$-minimization is still the same as that in
Theorem
\ref{thm:wl1}. We characterize the $\ell_p$-minimization ($p \in (0,1)$) case 
in Theorem \ref{thm:wb} and the $\ell_0$-minimization case in
Theorem \ref{thm:wl0}. 

\begin{theorem}\label{thm:wb}
Given any $p \in (0,1)$, for all $\bfx \in \mathcal{R}^n$ on some
support $T$ with some fixed sign pattern, $\bfx$ is always the
unique solution to $\ell_p$-minimization problem (\ref{eqn:lp}), if
and only if the following condition holds:
\begin{equation}\label{eqn:wlp}
\|B_{T^-}\bfz\|_p^p \leq \|B_{T^c}\bfz\|_p^p
\end{equation}
for all non-zero $\bfz \in \mathcal{R}^{n-m}$ where $T^-=\{i \in T:
B_i\bfz
x_i <0\}$; moreover, if 
$B_{T^+}\bfz=\bm0$ where $T^+=\{i \in T: B_i\bfz x_i \geq 0\}$, it
further holds that
\begin{equation}\label{eqn:wlp2}
\|B_{T^-}\bfz\|_p^p<\|B_{T^c}\bfz\|_p^p.
\end{equation}
\end{theorem}

\begin{proof}
Necessary part. Suppose the condition fails for some $\bfz$, then
there are two cases: either $B_{T^+}\bfz=\bm0$ or $B_{T^+}\bfz \neq
\bm0$.

First consider the case $B_{T^+}\bfz=\bm0$, then we have
$\|B_{T^-}\bfz\|_p^p \geq \|B_{T^c}\bfz\|_p^p$. Define a vector
$\bfx$ as follows. Let $x_i =0$ for every $i$ in $T^c$, let
$x_i=-B_i\bfz$ for every $i$ in $T^-$. Let $x_i$ be any value with
the fixed sign for every $i$ in $T^+$. Then according to the
definition of $\bfx$, we have
\begin{eqnarray*}
& &\|\bfx+B\bfz\|_p^p \\&=& \|\bfx_{T^{-}}+B_{T^-}\bfz\|_p^p+ \|\bfx_{T^{+}}+B_{T^+}\bfz\|_p^p+\|B_{T^c}\bfz\|_p^p\\
&=&0+ \|\bfx_{T^{+}}\|_p^p+\|B_{T^c}\bfz\|_p^p\\
&=& \|\bfx\|_p^p-\|\bfx_{T^{-}}\|_p^p +\|B_{T^c}\bfz\|_p^p\\
&=& \|\bfx\|_p^p-\|B_{T^-}\bfz\|_p^p +\|B_{T^c}\bfz\|_p^p\\
&\leq&  \|\bfx\|_p^p.
\end{eqnarray*}
Since $\|\bfx+B\bfz\|_p^p  \leq \|\bfx\|_p^p$, (\ref{eqn:lp}) cannot
successfully recover $\bfx$, which is a contradiction.

Secondly, consider the case $B_{T^+}\bfz \neq \bm0$. Then
$\|B_{T^-}\bfz\|_p^p > \|B_{T^c}\bfz\|_p^p$.
Let $\delta=\|B_{T^-}\bfz\|_p^p- \|B_{T^c}\bfz\|_p^p>0$. Define a
vector $\bfx$ as follows. Let $x_i =0$ for every $i$ in $T^c$, let
$x_i=-B_i\bfz$ for every $i$ in $T^-$. For every $i$ in $T^+$, since
$p \in (0,1)$, we can pick $x_i$ with $|x_i|$ large enough such that
$\|\bfx_{T^+}+B_{T^+}\bfz\|_p^p - \|\bfx_{T^+}\|_p^p<
\frac{\delta}{2}$. Then
\begin{eqnarray*}
\|\bfx+B\bfz\|_p^p &=& 0+ \|\bfx_{T^{+}}+B_{T^+}\bfz\|_p^p+\|B_{T^c}\bfz\|_p^p\\
&<& \|\bfx_{T^{+}}\|_p^p + \frac{\delta}{2}+\|B_{T^c}\bfz\|_p^p\\
&=& \|\bfx_{T^{+}}\|_p^p + \frac{\delta}{2}+\|B_{T^-}\bfz\|_p^p-\delta\\
&=&  \|\bfx\|_p^p - \frac{\delta}{2}.
\end{eqnarray*}
Thus $\|\bfx+B\bfz \|_p^p < \|\bfx\|_p^p$, $\bfx$ is not a solution
to (\ref{eqn:lp}), which is also a contradiction.

Sufficient part. Assume the null space condition holds, then for any
$\bfx$ on support $T$ with fixed signs, and any non-zero $\bfz \in
\mathcal{R}^{n-m}$, we have 
 \noindent
\begin{eqnarray}
\hspace{-.35in}&&\|\bfx+B\bfz \|_p^p  \nonumber\\
\hspace{-.35in}&= & \|\bfx_{T^+}+B_{T^+}\bfz\|_p^p +\|\bfx_{T^-}+B_{T^-}\bfz\|_p^p+ \|B_{T^c}\bfz\|_p^p \nonumber\\
\hspace{-.35in}& \geq & \|\bfx_{T^+}+B_{T^+}\bfz\|_p^p
+\|\bfx_{T^-}\|_p^p-\|B_{T^-}\bfz\|_p^p+\|B_{T^c}\bfz\|_p^p,
\label{eqn:xbz}
\end{eqnarray}
where the inequality follows from the triangular property that
$|\bfx_i+B_i\bfz|^p \geq |\bfx_i|^p-|B_i\bfz|^p$ holds for all $i$
and all $p \in (0,1)$.

 If $B_{T^+}\bfz \neq \bm0$, then
$\|\bfx_{T^+}+B_{T^+}\bfz\|_p^p>\|\bfx_{T^+}\|_p^p$ since $B_i\bfz
\neq 0$ for some $i$, and $B_i \bfz$ and $x_i$ have the same sign.
Since we also have $\|B_{T^-}\bfz\|_p^p \leq \|B_{T^c}\bfz\|_p^p$,
therefore (\ref{eqn:xbz})$>\|\bfx\|_p^p$. If $B_{T^+}\bfz =\bm0$,
then $\|B_{T^-}\bfz\|_p^p <\|B_{T^c}\bfz\|_p^p$ from assumption,
therefore we also have (\ref{eqn:xbz})$>\|\bfx\|_p^p$. Thus,
$\|\bfx+B\bfz \|_p^p>\|\bfx\|_p^p$ for all non-zero $\bfz \in
\mathcal{R}^{n-m}$, then $\bfx$ is the solution to (\ref{eqn:lp}).
%
\end{proof}

Similarly, the null space condition for the weak recovery of
$\ell_0$-minimization is as follows, we skip its proof as it is
similar to that of Theorem \ref{thm:wb}.
\begin{theorem}\label{thm:wl0}
For all $\bfx \in \mathcal{R}^n$ on one support $T$ with the same
sign pattern, $\bfx$ is always the unique solution to
$\ell_0$-minimization problem (\ref{eqn:l0}), if and only if
\begin{equation}\label{eqn:wl0}
\|B_{T^-}\bfz\|_0 <\|B_{T^c}\bfz\|_0
\end{equation}
for all non-zero $\bfz \in \mathcal{R}^{n-m}$ where $T^-=\{i \in T:
B_i\bfz x_i <0\}$.
\end{theorem}

For the strong recovery, the null space conditions of
$\ell_1$-minimization and $\ell_p$-minimization ($0\leq p<1$) share
the same form (\ref{eqn:slp}), and if (\ref{eqn:slp}) holds for some
$p \leq 1$, it also holds for all $q \in [0,p]$. However, for
recovery of sparse vectors on one support with one sign pattern,
from Theorem \ref{thm:wl1}, \ref{thm:wb} and \ref{thm:wl0}, we know
that although the conditions of $\ell_p$-minimization ($0<p<1$) and
$\ell_0$-minimization share a similar form in (\ref{eqn:wlp}),
(\ref{eqn:wlp2}) and (\ref{eqn:wl0}), the condition of
$\ell_1$-minimization has a very different form in (\ref{eqn:wl1}).
Moreover, if (\ref{eqn:wlp}) holds for some $p \in (0,1)$, it does
not necessarily hold for some $q
\in (0, p)$. 
Therefore the way that the performance of weak recovery changes over
$p$ may be quite different from the way that the performance of
strong recovery changes over $p$. Moreover, the performance of weak
recovery of $\ell_1$ may be significantly different from that of
$\ell_p$-minimization for $ p\in (0,1)$. We will further discuss
this issue.

\subsection{The solution of $\ell_1$-minimization can be sparser than
that of $\ell_p$-minimization ($p\in (0,1)$)}\label{sec:example}

$\ell_p$-minimization ($p \in (0,1)$) may not perform as well as
$\ell_1$-minimization in some cases, 
for example in the weak recovery which we will discuss in Section
\ref{sec:limit} and Section \ref{sec:finite}. Here we employ a
numerical example to illustrate that in certain cases
$\ell_1$-minimization can recover the sparse vector while
$\ell_p$-minimization ($p\in (0,1)$) cannot, and the solution of
$\ell_p$-minimization is denser than the original sparse vector.

\noindent \textbf{Example 1.} $\ell_p$-minimization returns a denser
solution than $\ell_1$-minimization.

Let the measurement matrix $A$ be a $(6k-1)\times 6k$ matrix with
$\bm\beta \in \mathcal{R}^{6k}$ as a basis of its null space, and
$\beta_i=1$ for all $i\in \{1,...,k\}$, $\beta_i=-1$ for all $i \in
\{k+1,...,2k\}$, and $\beta_i=1/64$ for all $i \in \{2k+1,...,6k\}$.
According to Theorem \ref{thm:slp}, one can calculate that
$\ell_1$-minimization can recover all the $(\lceil
\frac{33}{32}k\rceil-1)$-sparse vectors in $\mathcal{R}^{6k}$, and
$\ell_{0.5}$-minimization can recover all the
$(\lceil\frac{5}{4}k\rceil-1)$-sparse vectors in $\mathcal{R}^{6k}$.
Therefore, in terms of strong recovery, $\ell_{0.5}$-minimization
has a better performance than $\ell_1$-minimization as it can
recover all the vectors up to a higher sparsity.

Now consider the ``weak'' recovery as to recover all the nonnegative
vectors on support $T=\{1,...,2k\}$. According to Theorem
\ref{thm:wl1} and Theorem \ref{thm:wb}, one can check that
$\ell_1$-minimization can indeed recover all the nonnegative vectors
on support $T$, however, $\ell_{0.5}$-minimization fails to recover
some vectors in this case. For example, consider a $2k$-sparse
vector $\bfx^*$ with $x^*_i=9$ for all $i\in \{1,...,k\}$, $x^*_i=1$
for all $i\in\{k+1,...,2k\}$, and $x^*_i=0$ for all
$i\in\{2k+1,...,6k\}$. One can check that among all the vectors
$\bfx=\bfx^*+h\bm\beta$, $\forall h \in \mathcal{R}$, which are the
solutions to $A\bfx=A\bfx^*$, $\bfx^*$ has the least $\ell_1$ norm,
therefore $\bfx^*$ is the solution to (\ref{eqn:l1}) and can be
successfully recovered via $\ell_1$-minimization.  Now consider
$\ell_{0.5}$-minimization, we have $\|\bfx^*\|_{0.5}^{0.5}=4k$.
Consider the nonnegative $5k$-sparse vector $\bfx'=\bfx^*+\bm\beta$
with $x'_i=10$ for all $i\in\{1,...,k\}$, $x'_i=0$ for all
$i\in\{k+1,...,2k\}$, and $x'_i=1/64$ for all $i\in\{2k+1,...,6k\}$.
We have $A\bfx'=A\bfx^*$, and one can check that
$\|\bfx'\|_{0.5}^{0.5}=(\sqrt{10}+0.5)k<\|\bfx^*\|_{0.5}^{0.5}$ for
all $k \geq 2$. Moreover, with a little calculation one can prove
that $\bfx'$ is indeed the solution to (\ref{eqn:lp}). Thus, the
solution of $\ell_{0.5}$-minimization is a 
$5k$-sparse vector although the original vector $\bfx^*$ is only
$2k$-sparse. Therefore 
$\ell_{0.5}$-minimization fails to recover some nonnegative
$2k$-sparse vector $\bfx^*$ while $\bfx^*$ is the solution to
$\ell_1$-minimization, 
and the
solution of $\ell_{0.5}$-minimization is denser than the original
vector $\bfx^*$.

\section{Recovery thresholds when $\lim_{n \rightarrow \infty}\frac{m}{n} \rightarrow
1$}\label{sec:limit}

In this paper we focus on the case that each entry of the
measurement matrix $A$ is drawn from standard Gaussian distribution.
Since $A$ has i.i.d. $\mathcal{N}(0,1)$ entries, the null space of
$A$ is rotationally invariant, thus there exists a basis $B^{n
\times (n-m)}$ of the null space of $A$ such that $AB=\bf{0}$ and
$B$ has i.i.d. $\mathcal{N}(0,1)$ entries, please refer to
\cite{CaT06}\cite{XH10} for details.

We first focus on the case that $\alpha=\frac{m}{n}\rightarrow 1$
and provide recovery thresholds of $\ell_p$-minimization for every
$p\in [0,1]$. we consider two types of thresholds: one in the
\emph{strong} sense as we require $\ell_p$-minimization to recover
\textit{all} $\rho n$-sparse vectors (Section \ref{sec:sbd}), one in
the \emph{weak} sense as we only require $\ell_p$-minimization to
recover \textit{all the vectors on a certain support with a certain
sign pattern} (Section \ref{sec:wbd}). We call it a threshold as for
any sparsity below that threshold, $\ell_p$-minimization can recover
all the sparse vectors either in the strong sense or the weak sense,
and for any sparsity above that threshold, $\ell_p$-minimization
fails to recover some sparse vector. These thresholds can be viewed
as the limiting behavior of $\ell_p$-minimization, since for any
constant $\alpha<1$, the recovery thresholds of
$\ell_p$-minimization would be no greater than the ones provided
here.

\subsection{Strong Recovery}\label{sec:sbd}
In this section, for given $p$, when $\alpha \rightarrow 1$, we
shall provide a threshold
$\rho^*(p)$ for \textit{strong recovery} such that for any $\rho<\rho^*(p)$, 
 $\ell_p$-minimization (\ref{eqn:lp}) can recover \textit{all} $\rho n$-sparse vectors
$\bfx$ with overwhelming probability. Our technique here stems from
\cite{DMT07}, which only focuses on the strong recovery of
$\ell_1$-minimization.

We have already discussed in Section \ref{sec:null} that the
performance of $\ell_q$-minimization should be no worse than
$\ell_p$-minimization for strong recovery when $0\leq q<p\leq 1$.
Although there are results about bound of the sparsity below which
$\ell_p$-minimization can recover all the sparse vectors, no
existing result has explicitly calculated the recovery threshold of
$\ell_p$-minimization  for $p<1$ which differentiates the success
and failure of $\ell_p$-minimization. To this end, we will first
define $\rho^*(p)$ in the following lemma, and then prove that
$\rho^*(p)$ is indeed the threshold of strong recovery in later
part.
\begin{lemma}
Let $X_1$, $X_2$,...,$X_n$ be i.i.d $\mathcal{N}(0,1)$ random
variables and let $Y_1$, $Y_2$,...,$Y_n$ be the sorted ordering (in
non-increasing order) of $|X_1|^p$, $|X_2|^p$,...,$|X_n|^p$ for some
$p \in (0,1]$. For a $\rho
>0$, define $S_\rho$ as $\sum \limits_{i=1}^{\lceil \rho n \rceil}
Y_i$. Let $S$ denote $E[S_1]$, the expected value of $S_1$. Then
there exists a constant $\rho^*(p)$ such that $\lim \limits_{n
\rightarrow \infty} \frac{E[S_{\rho^*}]}{S}=\frac{1}{2}$.
\end{lemma}
\begin{proof}
Let $X \sim \mathcal{N}(0,1)$ and let $Z=|X|$. Let $f(z)$ and $F(z)$
denote the p.d.f. and c.d.f. of $Z$ respectively. Then
\begin{eqnarray}
f(z)&=&\sqrt{2/\pi}e^{-\frac{1}{2}z^2},  \quad \textrm{if }
z\geq 0,  \nonumber \\
&=&0, \quad \textrm{if }  z<0.\label{eqn:pdf}
\end{eqnarray}
\begin{eqnarray}
F(z)&=& \textrm{erf}(z/\sqrt{2})=\int_{0}^{z}
\sqrt{2/\pi}e^{-\frac{1}{2}x^2}dx, \quad \textrm{if }
z\geq 0,  \nonumber \\
&=&0,  \quad \textrm{if } z<0. \label{eqn:cdf}
\end{eqnarray}

 Define $g(t)= \int_{t}^ \infty z^p f(z)dz$. $g$ is continuous
and decreasing in $[0, \infty]$, and $g(0)=E[Z^p]=\frac{S}{n}$,
$\lim _{t \rightarrow \infty} g(t)=0$. Then there exists $z^*$ such
that $g(z^*)=\frac{g(0)}{2}$, i.e.
\begin{equation} \label{eqn:zstar}
\int_0^{z^*}x^pf(x)dx-\int_{z^*}^\infty x^pf(x)dx=0.
\end{equation}

Define
\begin{equation}\label{eqn:rhostar}
\rho^*=1-F(z^*).
\end{equation}
We claim $\rho^*$ has the desired property.

Let $T_t=\sum_{i:Y_i \geq t^p}Y_i$. Then $E[T_{z^*}]=ng(z^*)$. Since
$E[|T_{z^*}-S_{\rho^*}|]$ is bounded by $O(\sqrt{n})$, and
$S=ng(0)$, thus $\lim _{n \rightarrow \infty}
\frac{E[S_{\rho^*}]}{S}=\frac{1}{2}.$

%
\end{proof}

\begin{prop}\label{prop:rho}
The function $\rho^*(p)$ is strictly decreasing 
in $p$ on $(0,1]$.
\end{prop}
\begin{proof}
From the definition of $z^*$ in (\ref{eqn:zstar}), we have
\begin{equation} \label{eqn:zp}
H(z^*,p):=\int_0^{z^*}x^pf(x)dx -\int_{z^*}^\infty x^pf(x)dx=0,
\end{equation}
  where $f(\cdot)$
and $F(\cdot)$ are 
defined in (\ref{eqn:pdf}) and (\ref{eqn:cdf}). From the Implicit
Function Theorem,
\begin{equation}\nonumber
\frac{dz^*}{dp}=-\frac{\frac{\partial H}{\partial p}}{\frac{\partial
H}{\partial z^{*}}}=-\frac{\int_0^{z^*} x^p(\ln x)f(x)dx
-\int_{z^*}^\infty x^p(\ln x) f(x)dx}{2z^{*p}f(z^*)}.
\end{equation}
From (\ref{eqn:rhostar}), we have $\frac{d\rho^*}{dz^*}=-f(z^*)$.
From the chain rule, we know
$\frac{d\rho^*}{dp}=\frac{d\rho^*}{dz^*}\frac{dz^*}{dp}$, thus
\begin{equation}\label{eqn:drhodp}
\frac{d\rho^*}{dp}
=\frac{\int_0^{z^*} x^p(\ln x)f(x)dx -\int_{z^*}^\infty x^p(\ln x)
f(x)dx}{2z^{*p}}
\end{equation}

Note that
\begin{eqnarray}
\int_0^{z^*} x^p(\ln x)f(x)dx &<&\int_0^{z^*} x^p(\ln z^*)f(x)dx \nonumber\\
&=&\int_{z^*}^\infty x^p (\ln z^*)f(x)dx \nonumber\\
&<&\int_{z^*}^\infty x^p(\ln x) f(x)dx, \label{eqn:zpp}
\end{eqnarray}
where the equality follows from (\ref{eqn:zp}). Then the numerator
of (\ref{eqn:drhodp}) is less than 0 from (\ref{eqn:zpp}), thus
$\frac{d\rho^*}{dp}<0$.

\end{proof}

\begin{figure}[t]
      \centering
      \includegraphics[scale=0.5]{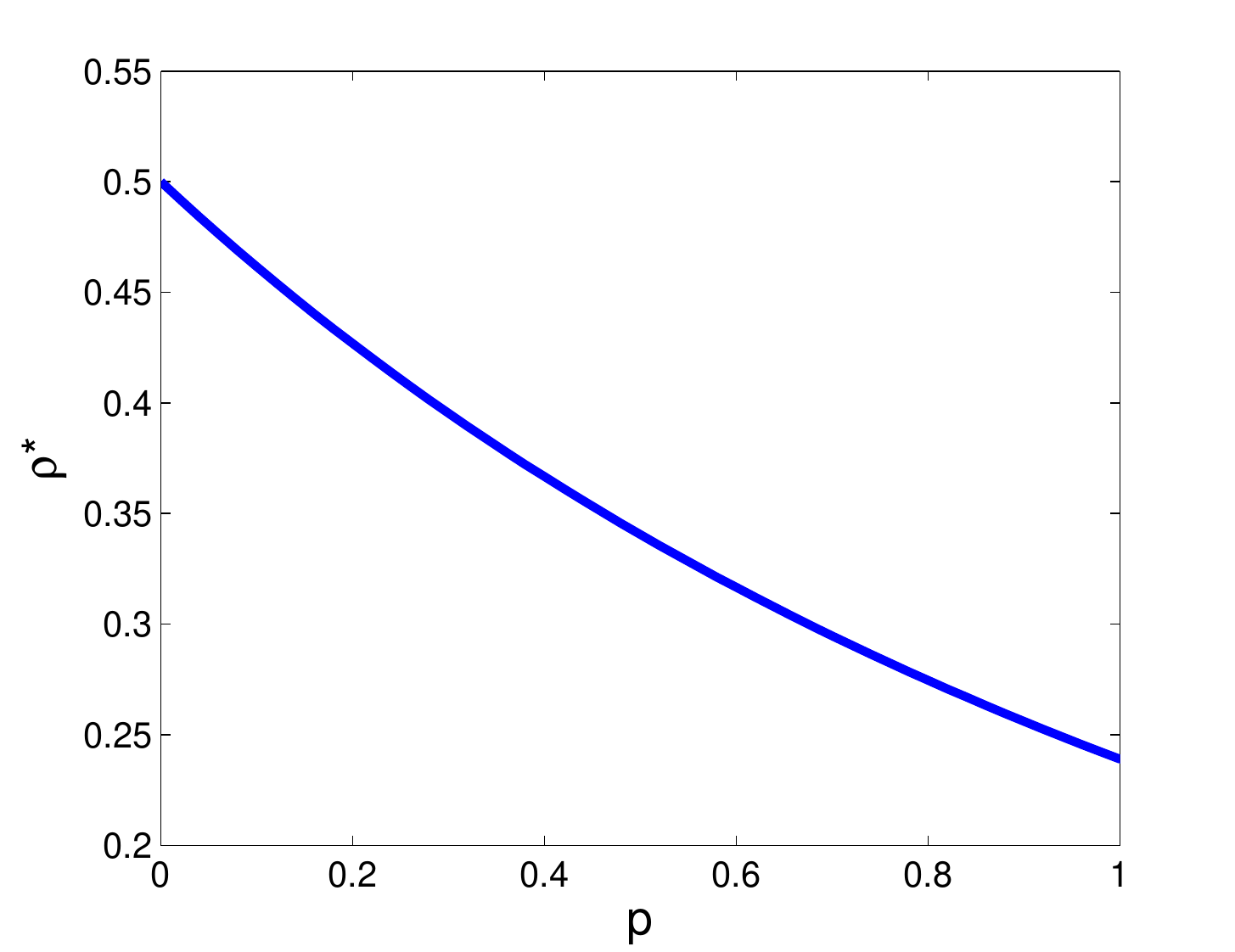}
 \caption{Threshold $\rho^*$ of successful recovery with $\ell_p$-minimization}
   \label{fig:rho}
\end{figure}

We plot $\rho^*$ against $p$ numerically in Fig. \ref{fig:rho}.
$\rho^*(p)$ goes to $\frac{1}{2}$ as $p$ tends to zero. Note that
$\rho^*(1)=0.239...$, which coincides with the result in
\cite{DMT07}.

Now we proceed to prove that $\rho^*$ is the threshold of successful
recovery with $\ell_p$ minimization for $p$ in $(0,1]$. First we
state the concentration property of $S_\rho$ in the following lemma.
\begin{lemma}\label{lemma:srho}
For any $p \in (0,1]$, let $X_1$,...,$X_n$, $Y_1$,...,$Y_n$,
$S_\rho$ and $S$ be as above. For any $\rho>0$ and any $\delta>0$,
there exists a constant $c_1>0$ such that when $n$ is large enough,
with probability at least $1-2e^{-c_1n}$, $|S_\rho-E[S_\rho]| \leq
\delta S$.
\end{lemma}

\begin{proof}
Let $\bfX=[X_1,...,X_n]^T$. If two vectors $\bfX$ and $\bfX'$ only
differ in co-ordinate $i$, then for any $p$,
$|S_\rho(\bfX)-S_\rho(\bfX')| \leq ||X_i|^p-|X'_i|^p|$. Thus for any
$\bfX$ and $\bfX'$,
\begin{equation}\nonumber
|S_\rho(\bfX)-S_\rho(\bfX')| \leq \sum_{i: X_i \neq X'_i}
\big||X_i|^p-|X'_i|^p\big|.
\end{equation}
Since $\big||X_i|^p-|X'_i|^p\big| \leq |X_i-X'_i|^p$ for all $p \in
(0,1]$,
\begin{equation}\label{eqn:srho}
|S_\rho(\bfX)-S_\rho(\bfX')|  \leq \sum_{i} |X_i-X'_i|^p.
\end{equation}

 From
 the isoperimetric inequality for the Gaussian measure \cite{Ledoux01}, for any set
 $A$ with measure at least a half, the set $A_t=\{\bfx \in
 \mathcal{R}^n: d(\bfx,A) \leq t\}$ has measure at least
 $1-e^{-t^2/2}$, where $d(\bfx,A)= \inf_{\bfy \in A} \|\bfx-\bfy\|_2$.
  Let $M_\rho$ be the median value of $S_\rho=S_\rho (\bm X)$. Define set $A=\{\bfx \in \mathcal{R}^n: S_\rho(\bfx) \leq M_\rho\}$, then
\begin{equation}\nonumber
P(d(\bfx,A) \leq t)\geq 1-e^{-t^2/2}.
\end{equation}
We claim that $d(\bfx,A) \leq t$ implies that $S_\rho(\bfx) \leq
M_\rho+n^{(1-p/2)}t^p$. If $\bfx \in A$, then $S_\rho(\bfx) \leq
M_\rho$, thus the claim holds as $n^{1-p/2}t^p$ is nonnegative. If
$\bfx \notin A$, then there exists $\bfx' \in A$ such that
$\|\bfx-\bfx'\|_2 \leq t$. Let $u_i=1$ for all $i$ and let $v_i
=|x_i-x'_i|^p$. From H\"older's inequality,
\begin{eqnarray}
\sum_{i} |x_i-x'_i|^p  
& \leq & \left( \sum_{i}|u_i|^{2/(2-p)}\right)^{1-p/2}  \left( \sum_{i} |v_i|^{2/p}\right)^{p/2} \nonumber \\
&\leq & n^{(1-p/2)} (t^2)^{p/2} = n^{(1-p/2)}t^p \label{eqn:holder}
\end{eqnarray}

From (\ref{eqn:srho}) and (\ref{eqn:holder}),
$|S_\rho(\bfx)-S_\rho(\bfx')| \leq n^{(1-p/2)}t^p$. Since $ \bfx
\notin A$ and $\bfx' \in A$, then $S_\rho(\bfx) > M_\rho \geq
S_\rho(\bfx')$. Thus $S_\rho(\bfx) \leq M_\rho+n^{(1-p/2)}t^p$,
which verifies our claim. Then
\begin{equation}\label{eqn:srholb}
P(S_\rho(\bfx) \leq M_\rho+n^{(1-p/2)}t^p)\geq P(d(\bfx,A) \leq
t)\geq 1-e^{-t^2/2}.
\end{equation}
Similarly,
\begin{equation}\label{eqn:srhoub}
P(S_\rho(\bfx) \geq M_\rho-n^{(1-p/2)}t^p)\geq 1-e^{-t^2/2}.
\end{equation}
Combining (\ref{eqn:srholb}) and (\ref{eqn:srhoub}),
\begin{equation}\label{eqn:srhomrho}
P(|S_\rho(x)-M_\rho| \geq n^{(1-p/2)}t^p)\leq 2e^{-t^2/2}.
\end{equation}

The difference of $E[S_{\rho}]$ and $M_{\rho}$ can be bounded as
follows,
\begin{eqnarray*}
|E[S_\rho]-M_\rho|
&\leq& E[|S_\rho-M_\rho|]\\
&=&\int_0^\infty P(|S_\rho(x)-M_\rho| \geq y) dy\\
&\leq &\int_0^\infty 2e^{-\frac{1}{2}y^{\frac{2}{p}}n^{(1-\frac{2}{p})}} dy\\
&=&n^{(1-\frac{p}{2})}\int_0^\infty 2e^{-\frac{1}{2}s^{\frac{2}{p}}}ds\\
\end{eqnarray*}

Note that $c:=\int_0^\infty 2e^{-\frac{1}{2}s^{(2/p)}}ds$ is a
finite constant for all $p \in (0,1]$. As $p>0$ and $S=nE[|x_i|^p]$,
thus for any $\delta >0$, $cn^{(1-\frac{p}{2})} <\frac{\delta}{2}S$
when $n$ is large enough.

Let $t=\left(\frac{1}{2}\delta S
n^{(\frac{p}{2}-1)}\right)^\frac{1}{p}=(\frac{1}{2}\delta
E[|x_i|^p])^{\frac{1}{p}}\sqrt{n}$, from (\ref{eqn:srhomrho}) with
probability at least $1- 2e^{-\frac{1}{2}(\frac{1}{2}\delta
E[|x_i|^p])^{\frac{2}{p}}n}$, $|S_\rho-M_\rho| < \frac{1}{2}\delta
S$. Thus $|S_\rho-E[S_\rho]| \leq
|S_\rho-M_\rho|+|M_\rho-E[S_\rho]|< \delta S$ with probability at
least $1-2e^{-c_1n}$ for some constant $c_1$.
\end{proof}
\begin{cor}\label{cor:srho}
For any $\rho < \rho^*$, there exists a $\delta>0$ and a constant
$c_2>0$ such that when $n$ is large enough, with probability at
least $1-2e^{-c_2n}$, $S_\rho \leq (\frac{1}{2}-\delta)S$.
\end{cor}
\begin{proof}
When $\rho < \rho^*$,
\begin{eqnarray*}
E[S_\rho]&=&E[S_{\rho^*}]-\sum \limits_{i=\lceil \rho n\rceil+1}^{\lceil \rho^*n \rceil} E[|X_i|^p]\\
& \leq & E[S_{\rho^*}]-(\lceil \rho^*n \rceil-\lceil \rho n\rceil)
E[|X_i|^p]
\end{eqnarray*}
Then $E[S_\rho]/S \leq \frac{1}{2}-2\delta$ for a suitable $\delta$
as $S=nE[|X_i|^p]$. The result follows by combining the above with
Lemma \ref{lemma:srho}.
\end{proof}

\begin{cor}\label{cor:s1}
For any $\epsilon >0$, there exists a constant $c_3>0$ such that
when $n$ is large enough, with probability at least $1-2e^{-c_3n}$,
it holds that $(1-\epsilon)S \leq S_1\leq (1+\epsilon)S$.
\end{cor}

The above two corollaries indicate that with overwhelming
probability the sum of the largest $\lceil \rho n \rceil $ terms of
$Y_i$'s is less than half of the total sum $S_1$ if $\rho< \rho^*$.
The following lemma extends the result to every vector $B\bfz$ where
matrix $B^{n \times (n-m)}$ has i.i.d. Gaussian entries and $\bfz$
is any non-zero vector in $\mathcal{R}^{n-m}$.
\begin{lemma}\label{lemma:slp}
For any $0<p \leq 1$, given any $\rho < \rho^*(p)$, there exist
constants $0<c_4<1$, $c_5>0$, $\delta >0$ such that when
$\alpha=\frac{m}{n}> c_4$ and $n$ is large enough, with probability
at least $1-e^{-c_5n}$, an $n \times (n-m)$ matrix $B$ with i.i.d.
$\mathcal{N}(0,1)$ entries has the following property: for every
non-zero $\bfz \in \mathcal{R}^{n-m}$ and every subset $T \subseteq
\{1,...,n\}$ with $|T| \leq \rho n$, $ \|B_{T^c}\bfz\|_p^p -
\|B_T\bfz\|_p^p \geq
\delta S \|\bfz\|_2^p$. 
\end{lemma}

\begin{proof}
For any given $\gamma>0$, there exists a $\gamma$-net $\Sigma$ in
$\mathcal{R}^{n-m}$ of cardinality less than
$(1+\frac{2}{\gamma})^{n-m}$(\cite{Ledoux01}). A $\gamma$-net
$\Sigma$ is a set of points in $\mathcal{R}^{n-m}$ such that
$\|\bfv^k\|_2=1$ for all $\bfv^k$ in $\Sigma$ and for any $\bfz\in
\mathcal{R}^{n-m}$ with $\|\bfz\|_2=1$, there exists some
$\bfv^k$ such that $\|\bfz-\bfv^k\|_2 \leq \gamma$. 

Since $B$ has i.i.d $\mathcal{N}(0,1)$ entries, then $B\bfv^k$ has
$n$ i.i.d. $\mathcal{N}(0,1)$ entries for every $\bfv^k$. From
Corollary \ref{cor:srho} and \ref{cor:s1}, we know that given any
$\rho<\rho^*$, for some $\delta>0$ and for every $\epsilon
>0$, there exists $c_2>0$ and $c_3$ such that with probability at least $1-2e^{-c_2n}-2e^{-c_3n}$, we have
\begin{equation}\label{prop:srho}
S_\rho(Av^k) \leq (\frac{1}{2}-\delta)S
\end{equation} and
\begin{equation}\label{prop:s1}
(1-\epsilon)S \leq S_1(Av^k)\leq (1+\epsilon)S
\end{equation}
both hold for \textit{a} vector $\bfv^k$ in $\Sigma$. Then applying
union bound, we know that (\ref{prop:srho}) and (\ref{prop:s1}) hold
for \textit{all} vectors in $\Sigma$ with probability at least
\begin{equation}\label{eqn:probsuc}
1-(1+2/\gamma)^{n-m}(2e^{-c_2n}+2e^{-c_3n}).
\end{equation}
Let $\alpha=m/n$, then as long as
$\alpha>c_4:=1-\frac{\min(c_2,c_3)}{\ln (1+2/\gamma)}$, then
(\ref{eqn:probsuc})$\geq 1-e^{-c_5n}$ for some constant $c_5>0$.

 For any $\bfz$ such that $\|\bfz\|_2=1$, there exists $\bfv_0$ in $\Sigma$ such
 that $\|\bfz-\bfv_0\|_2\triangleq \gamma_1 \leq \gamma$. Let $\bfz_1$ denote $\bfz-\bfv_0$,
 then $\|\bfz_1-\gamma_1\bfv_1\|_2 \triangleq \gamma_2 \leq \gamma_1 \gamma \leq \gamma^2$ for
 some $\bfv_1$ in $\Sigma$. Repeating this process, we have
\begin{equation}\label{eqn:expansion}
 \bfz=\sum_{j\geq 0} \gamma_j \bfv_j
\end{equation}
where $\gamma_0=1$, $\gamma_j \leq \gamma^j$ and $\bfv_j \in
\Sigma$. Thus for any $\bfz \in \mathcal{R}^{n-m}$, we have
 $\bfz=\|\bfz\|_2\sum_{j\geq 0} \gamma_j \bfv_j$.

For any index set $T$ with $|T| \leq \rho n$,
\begin{eqnarray*}
\|B_T\bfz\|_p^p &=& \|\bfz\|_2^p \|\sum \limits_{j \geq 0} \gamma_j B_T\bfv_j\|_p^p \\
& \leq & \|\bfz\|_2^p  \sum \limits_{j \geq 0} \gamma^{jp} \|B_T\bfv_j\|_p^p \\
& \leq & S\|\bfz\|_2^p \frac{1-2\delta}{2(1-\gamma^p)},
\end{eqnarray*}

\begin{eqnarray*}
 \|B\bfz\|_p^p &=& \|\bfz\|_2^p \|\sum \limits_{j \geq 0} \gamma_{j} B\bfv_j\|_p^p \\
& \geq & \|\bfz\|_2^p (\|Bv_0\|_p^p- \sum \limits_{j \geq 1} \gamma_{j}^p \|B\bfv_j\|_p^p) \\
& \geq & \|\bfz\|_2^p (\|B\bfv_0\|_p^p- \sum \limits_{j \geq 1} \gamma^{jp} \|B\bfv_j\|_p^p) \\
& \geq & \|\bfz\|_2^p ((1-\epsilon)S- \sum \limits_{j \geq 1} \gamma^{jp} (1+\epsilon)S) \\
& \geq &  S\|\bfz\|_2^p \frac{1-2\gamma^p-\epsilon}{1-\gamma^p}
\end{eqnarray*}

Thus  $\|B_{T^c}\bfz\|_p^p - \|B_T\bfz\|_p^p \geq S \|\bfz\|_2^p
\frac{2\delta-2\gamma^p-\epsilon}{1-\gamma^p}$.
For a given $\delta$, we can pick $\gamma$ and $\epsilon$ small
enough such that $\|B_{T^c}\bfz\|_p^p -\|B_T\bfz\|_p^p \geq \delta S
\|\bfz\|_2^p$.
\end{proof}

We can now establish one main result regarding the threshold of
successful recovery via $\ell_p$-minimization.
\begin{theorem}
For any $0<p \leq 1$, given any $\rho < \rho^*(p)$, there exist
constants $0<c_4<1$, $c_5 >0$ such that when $\alpha>c_4$ and $n$ is
large enough, with probability at least $1-e^{-c_5n}$, an $m \times
n$ matrix $A$ with i.i.d. $\mathcal{N}(0,1)$ entries has the
following property: for every $\bfx \in \mathcal{R}^n$ with its
support $T$ satisfying $|T| \leq \rho n$, $\bfx$ is the unique
solution to the $\ell_p$-minimization problem (\ref{eqn:lp}).
\end{theorem}

\begin{proof}
Lemma \ref{lemma:slp} indicates that $\sum _{i \in T^c}
|(B\bfz)_i|^p - \sum _{i \in T} |(B\bfz)_i|^p \geq \delta S
\|z\|_2^p>0$ for every non-zero $z$, then from Theorem
\ref{thm:slp}, $\bfx$ is the unique solution to the
$\ell_p$-minimization problem (\ref{eqn:lp}).
\end{proof}

We remark here that $\rho^*$ is a sharp bound for successful
recovery. For any $\rho>\rho^*$, from Lemma \ref{lemma:srho}, with
overwhelming probability the sum of the largest $\lceil \rho
n\rceil$ terms of $|B_i\bfz|^p$'s is more than the half of the total
sum $S_1$, i.e. the null space condition stated in Theorem
\ref{thm:slp} for successful recovery via $\ell_p$-minimization
fails with overwhelming probability. Therefore,
$\ell_p$-minimization fails to recover some $\rho n$-sparse vector
with overwhelming probability.
Proposition \ref{prop:rho} implies that the threshold strictly
decreases as $p$ increases. The performance of
$\ell_{p_1}$-minimization is better than that of
$\ell_{p_2}$-minimization
for $0<p_1 <p_2 \leq 1$ 
as $\ell_{p_1}$-minimization can recover vectors up to a higher
sparsity.

\subsection{Weak Recovery}\label{sec:wbd}

We have demonstrated in Section \ref{sec:sbd} that the threshold for
strong recovery strictly decreases as $p$ increases from 0 to 1.
Here we provide a weak recovery threshold for all $p \in [0,1)$ when
$\alpha \rightarrow 1$. As we shall see, for weak recovery, the
threshold of $\ell_p$-minimization is the same for all $p \in
[0,1)$, and is lower than the threshold of $\ell_1$-minimization.

Recall that for successful weak recovery, $\ell_p$-minimization
should recover all the vectors on some fixed support with a fixed
sign pattern, and the equivalent null space characterization is
stated in Theorem \ref{thm:wb} and Theorem \ref{thm:wl0}.

We define $x^0=1$ for all $x \neq 0$, and $0^0=0$. 
To characterize the recovery threshold of $\ell_p$-minimization in
this case, we first state the following lemma,
\begin{lemma}\label{lemma:rhow}
Let $X_1$, $X_2$,...,$X_n$ be i.i.d. $\mathcal{N}(0,1)$ random
variables and $T$ be a set of indices with size $|T|=\rho n$ for
some $\rho>0$. Let $\bfx \in \mathcal{R}^n$ be any vector on support
$T$ with fixed sign pattern. For every $p \in [0,1)$, for every $
\epsilon
>0$, when $n$ is large enough,  with probability at least $1-e^{-c_6n}$ for
some constant $c_6>0$, the following two properties hold
simultaneously:
\begin{itemize}
\item $\frac{1}{2}\rho n (\mu-\epsilon) < \sum_{i \in T: X_ix_i <0} |X_i|^p < \frac{1}{2}\rho n (\mu+\epsilon)$ 
\item $ (1-\rho) n (\mu-\epsilon) < \sum_{i \in T^c} |X_i|^p < (1-\rho)
n (\mu+\epsilon)$.
\end{itemize}
where $\mu=E[|X|^p]$, $X \sim \mathcal{N}(0,1)$.
\end{lemma}

\begin{proof}
Define a random variable $s_i$ for each $i$ in $T$ that is equal to
1 if $X_ix_i <0$ and equal to 0 otherwise. Then $\sum_{i \in T:
X_ix_i <0} |X_i|^p=\sum_{i \in T} |X_i|^ps_i$. $
E[|X_i|^ps_i]=\frac{1}{2}\mu$ for every $i$ in $T$ as $X_i \sim
\mathcal{N}(0,1)$. From the Chernoff bound, for any $\epsilon >0$,
there exist $d_1>0$ and $d_2>0$ such that
\begin{itemize}
\item[] $P[\sum_{i \in T} |X_i|^px_i \leq \frac{1}{2}\rho n (\mu-\epsilon)] \leq e^{-d_1 n}$, 

\item[] $P[\sum_{i \in T} |X_i|^px_i \geq \frac{1}{2}\rho n (\mu+\epsilon)] \leq e^{-d_2 n}.$ 
\end{itemize}
Again from the Chernoff bound, there exist some constants $d_3>0$,
$d_4>0$ such that
\begin{itemize}
\item[]
$P[\sum_{i \in T^c} |X_i|^p \leq (1-\rho) n (\mu-\epsilon)] \leq e^{-d_3 n},$ 
\item[]
$P[\sum_{i \in T^c} |X_i|^p \geq (1-\rho) n (\mu+\epsilon)] \leq e^{-d_4 n}.$ 
\end{itemize}
By union bound, 
there exists some constant $c_6>0$ such that the two properties
stated in the lemma hold at the same time with probability at least
$1-e^{-c_6 n}$.

\end{proof}

Lemma \ref{lemma:rhow} implies that $\sum_{i \in T: X_ix_i <0}
|X_i|^p< \sum_{i \in T^c} |X_i|^p$ holds with high probability when
$|T|=\rho n < \frac{2}{3}n$. Applying the similar net argument in
Section \ref{sec:sbd}, we can extend the result to every vector
$B\bfz$ where matrix $B^{n \times (n-m)}$ has i.i.d. Gaussian
entries and $\bfz$ is any non-zero vector in $\mathcal{R}^{n-m}$.
Then we can establish the main result regarding the threshold of
successful recovery with $\ell_p$-minimization from vectors on one
support with the same sign pattern.
\begin{theorem}\label{thm:wth}
For any $p \in [0,1)$, given any $\rho < \rho^*_w:= \frac{2}{3}$,
there exist constants $c_7 \in (0,1)$, $c_8 >0$ such that when
$\alpha > c_7$ and $n$ is large enough, with probability at least
$1-e^{-c_8n}$, an $m \times n$ matrix $A$ with i.i.d.
$\mathcal{N}(0,1)$ entries has the following property: for every
vector $\bfx$ on some support $T$ satisfying $|T| \leq \rho m$ with
fixed sign pattern on $T$, $\bfx$ is the unique solution to the
$\ell_p$-minimization problem.
\end{theorem}

\begin{proof}
From Lemma \ref{lemma:rhow}, applying similar arguments in the proof
of Lemma \ref{lemma:slp}, we get that 
when $\alpha>c_7$ for some $0<c_7<1$ and $n$ is large enough, with
probability $1-e^{-c_8n}$ for some $c_8>0$,
\begin{itemize}
\item $\frac{1}{2}\rho n (\mu-\epsilon)<\sum_{i \in T: (B_i \bfv)x_i
<0} |B_i\bfv|^p < \frac{1}{2}\rho n (\mu+\epsilon)$
\item $(1-\rho) n (\mu-\epsilon) < \sum_{i \in T^c} |B_i\bfv|^p
< (1-\rho) n (\mu+\epsilon)$
\end{itemize}
hold for all the vectors $\bfv$ in a $\gamma$-net $\Sigma$ at the
same time. Let $\mathcal{S}$ be the unit sphere in
$\mathcal{R}^{n-m}$. Pick any $\bfz \in \mathcal{S}$, from
(\ref{eqn:expansion}) we have
 $\bfz=\sum_{j\geq 0} \gamma_j \bfv_j$,
where $\gamma_0=1$, $\bfv_j \in \Sigma$ for all $j$ and  $\gamma_j
\leq \gamma^j$.

Given $\bfz$, let $T^-=\{i \in T: B_i\bfz x_i <0\}$. For any $i$ in
$T^-$,
\begin{eqnarray}
|B_i\bfz|^p&=& \big|\sum_{j \geq 0} \gamma_j B_i\bfv_j\big|^p \nonumber \\
&=& \big|\sum_{j: (B_i\bfv_j)x_i<0 } \gamma_j B_i\bfv_j+\sum_{j: (B_i\bfv_j)x_i \geq 0 } \gamma_j B_i\bfv_j\big|^p  \nonumber\\
&\leq&   \big|\sum_{j: (B_i\bfv_j)x_i<0 } \gamma_j B_i\bfv_j\big|^p \nonumber\\
&\leq&  \sum_{j: (B_i\bfv_j)x_i < 0} \gamma^{jp} |B_i\bfv_j|^p
\nonumber \label{eqn:oneminus}
\end{eqnarray}
where the first inequality holds as $(B_i\bfz)x_i <0$. Then
\begin{eqnarray}
\|B_{T^-}\bfz\|_p^p&\leq & \sum \limits_{i \in T^-}  \sum \limits_{j: (B_i\bfv_j)x_i < 0} \gamma^{jp} |B_i\bfv_j|^p \nonumber\\
& \leq &  \sum \limits_{i \in T}  \sum \limits_{j: (B_i\bfv_j)x_i < 0} \gamma^{jp} |B_i\bfv_j|^p \nonumber\\
& = &   \sum \limits_{j\geq 0}  \gamma^{jp} \sum \limits_{i \in T:
(B_i\bfv_j)x_i < 0}  |B_i\bfv_j|^p \label{eqn:BT-}
\\
&<& \frac{1}{2(1-\gamma^p)}\rho n(\mu+\epsilon).\label{eqn:BT-2}
\end{eqnarray}

We also have
\begin{eqnarray}
&& \|B_{T^c}\bfz\|_p^p=   \|(\sum \limits_{j \geq 0} \gamma_{j} B_{T^c}\bfv_j)\|_p^p \nonumber \\
& \geq &  \|B_{T^c}\bfv_0\|_p^p- \sum \limits_{j \geq 1} \gamma^{jp} \|B_{T^c}\bfv_j\|_p^p \nonumber \\
& > &  (1-\rho)n(\mu-\epsilon)- \sum \limits_{j \geq 1} \gamma^{jp} (1-\rho)n(\mu+\epsilon) \nonumber\\
& \geq &  (1-\rho)n \frac{\mu-2\mu\gamma^p-\epsilon}{1-\gamma^p}.
\label{eqn:BTc}
\end{eqnarray}

Combining (\ref{eqn:BT-2}) and (\ref{eqn:BTc}), we have for every
$\bfz \in
\mathcal{S}$, 
$\|B_{T^c}\bfz\|_p^p- \|B_{T^-}\bfz\|_p^p >
\frac{n\mu}{1-\gamma^p}\big(1-\frac{3}{2}\rho-2\gamma^p(1-\rho)-\frac{\epsilon}{\mu}(1-\frac{\rho}{2})\big)$. %
Then for every non-zero $\bfz \in \mathcal{R}^{n-m}$, we have
$\|B_{T^c}\bfz\|_p^p- \|B_{T^-}\bfz\|_p^p >\|\bfz\|_2^p
\frac{n\mu}{1-\gamma^p}\big(1-\frac{3}{2}\rho-2\gamma^p(1-\rho)-\frac{\epsilon}{\mu}(1-\frac{\rho}{2})\big)$.
For any $\rho< \frac{2}{3}$, we can pick $\gamma$ and $\epsilon$
small enough such that the righthand side 
is positive. The result follows by applying Theorem \ref{thm:wb} and
Theorem \ref{thm:wl0}.

\end{proof}
We remark here that $\rho^*_w$ is a sharp bound for successful
recovery in this setup. For any $\rho>\rho^*_w$, from Lemma
\ref{lemma:rhow}, with overwhelming probability that $\sum_{i \in T:
X_ih_i <0} |X_i|^p> \sum_{i \in T^c} |X_i|^p$, then Theorem
\ref{thm:wb} and Theorem \ref{thm:wl0} indicate that the
$\ell_p$-minimization ($p\in [0,1)$) fails to recover some $\rho
n$-sparse vector $\bfx$ in this case. Note that for a random
Gaussian measurement matrix, from symmetry one can check that this
results does not depend on the specific choice of support and sign
pattern. In fact, Theorem \ref{thm:wth} holds for any fixed support
and any fixed sign pattern.

Surprisingly, the successful recovery threshold $\rho^*_w$ when we
only consider recovering vectors on one support with one sign
pattern 
is $\frac{2}{3}$ for all $p$ in $[0,1)$ and is strictly less than
the threshold for $p=1$, which is 1 (\cite{Donoho06}). Thus in this
case, $\ell_1$-minimization has better recovery performance than
$\ell_p$-minimization ($p\in[0,1)$) in terms of the sparsity
requirement for the sparse vector. If we view the ability to recover
all the vectors up to certain sparsity as the ``worst'' case
performance, and the ability to recovery all the sparse vectors on
one support with one sign pattern as the ``expected'' case
performance, then although worst case performance can be improved if
we apply $\ell_p$-minimization with a smaller $p$,
$\ell_1$-minimization in fact has the best expected case performance
for all $p\in [0,1]$.

 It
might be counterintuitive at first sight to see that the weak
threshold of $\ell_0$-minimization is less than that of
$\ell_1$-minimization, so let us take a moment to consider what the
result means. We choose recovering all nonnegative vectors on some
support $T$ ($|T|=\rho n$) for the weak recovery, the argument
follows for all the other supports and all the other sign patterns.
The results about weak recovery threshold indicate that for any
$\rho\in (2/3,1)$, when $n$ is sufficiently large and $\alpha
\rightarrow 1$, for a random Gaussian measurement matrix $A$,
$\ell_1$-minimization would recover all the nonnegative vectors on
some support $T$ ($|T|=\rho n$) with overwhelming probability, while
$\ell_0$-minimization would fail to recover some nonnegative vector
on $T$ with overwhelming probability according to Theorem
\ref{thm:wth}. This can happen when there exists a nonnegative
vector $\bfx$ on support $T$ and a vector $\bfx'$ on support $T'$
such that $|T'| \leq |T|$, and $A\bfx=A\bfx'$. Note that $\bfx'$
could have negative entries, or $T'$ may not be a subset of $T$.
Therefore, if $\bfx$ is the sparse vector we would like to recover
from $A\bfx$, $\ell_0$-minimization would fail since
$\|\bfx'\|_0\leq \|\bfx\|_0$. However, $\|\bfx\|_1 < \|\bfx'\|_1$
should hold since $\ell_1$-minimization can successfully return
$\bfx$ as its solution. Of course when $\bfx'$ is the sparse vector
we would like to recover, $\ell_1$-minimization would return $\bfx$
and fail to recover $\bfx'$. However, since $\ell_1$-minimization
would recover all the nonnegative vectors on
 $T$, then either $T' \nsubseteq T$ holds or $\bfx'$ has negative
entries. Therefore when we consider recovering nonnegative vectors
on $T$ for the weak recovery, $\bfx'$ is not taken into account, and
$\ell_1$-minimization works better than $\ell_0$-minimization.
Therefore, although the performance of $\ell_1$-minimization is not
as good as that of $\ell_p$-minimization ($p \in [0,1)$) in the
strong recovery which requires to recover all the vectors up to
certain sparsity, $\ell_1$-minimization can recover all the $\rho
n$-sparse ($\rho>2/3$) vectors on some support with some sign
pattern, while for $\ell_p$-minimization ($p \in [0,1)$), the size
of the largest support on which it can recover all the vectors with
one sign
pattern is no greater than $2n/3$. 
Thus, when we aim to recover all the vectors up to certain sparsity,
$\ell_p$-minimization is better for smaller $p$, however, when we
aim to recover all the vectors on one support with one sign pattern,
$\ell_1$-minimization may have a better performance.

\section{Recovery Bounds for Every $\lim_{n \rightarrow \infty} \frac{m}{n}<1$}\label{sec:finite}
We considered the limiting case that $\alpha \rightarrow 1$ in
Section \ref{sec:limit} and provided the limiting thresholds of
sparsity ratio for successful recovery via $\ell_p$-minimization
both in the strong sense and in the weak sense. Here we focus on the
case that $\alpha$ is given ($0<\alpha<1$). For any $\alpha$ and
$p$, we will provide a bound $\rho^*(\alpha, p)$ for strong recovery
and a bound $\rho^*_w(\alpha, p)$ for weak recovery such that
$\ell_p$-minimization can recover all the $\rho^*(\alpha,
p)n$-sparse
vectors 
with overwhelming probability, and recover all the
$\rho^*_w(\alpha,p)n$-sparse vectors on one support with one sign
pattern with overwhelming probability. Note that the thresholds we
provided in Section \ref{sec:limit} is tight in the sense that for
any $\rho>\rho^*$ in the strong recovery or any $\rho > \rho^*_w$ in
the weak recovery, with overwhelming probability
$\ell_p$-minimization would fail to recover some $\rho n$ sparse
vector. However, $\rho^*(\alpha, p)$ and $\rho^*_w(\alpha, p)$ we
provide in this section are lower bounds for the thresholds of
strong recovery and weak recovery respectively, and might not be tight in general.

\subsection{Strong Recovery}\label{sec:sbdfinite}
As discussed in Section \ref{sec:limit}, since $A$ has i.i.d.
$\mathcal{N}(0,1)$ entries, there exists a basis $B$ of the null
space of $A$ with i.i.d. $\mathcal{N}(0,1)$ entries. 
Let $\mathcal{S}$ be the unit sphere in $\mathcal{R}^{n-m}$. From
Theorem \ref{thm:slp} we know that in order to successfully recover
all the $\rho n$-sparse vectors via $\ell_p$-minimization,
$\|B_T \bfz\|_p^p < \frac{1}{2} \|B \bfz\|_p^p$ should hold for
every non-zero vector $\bfz \in \mathcal{R}^n$, and every set
$T\subset \{1,...,n\}$ with $|T| \leq \rho n$. We will first
establish a lower bound of $\|B \bfz\|_p^p$ for all $\bfz \in
\mathcal{S}$ with overwhelming probability in Lemma
\ref{lem:lambdamin}. Lemma \ref{lem:rhobound} establishes the fact
that for any given constant $c>0$, there always exists some $\rho>0$
such that $\|B_T \bfz\|_p^p \leq cn$ for all $\bfz \in \mathcal{S}$
and all $T$ with $|T|\leq \rho n$ with overwhelming probability.
Combining Lemma \ref{lem:lambdamin} and Lemma \ref{lem:rhobound} we
will establish a positive lower bound $\rho^*(\alpha,p)$ of sparsity
ratio for successful recovery for every $\alpha \in (0,1)$ and every
$p \in (0,1]$ in Theorem \ref{thm:boundalpha}.

\begin{lemma}\label{lem:lambdamin}
For any $\alpha$ and $p$, there exists a constant
$\lambda_{\min}(\alpha, p)>0$ and some constant $c_9>0$ 
such that with probability at least
$1-e^{-c_9n}$, for every $\bfz \in \mathcal{S}$, $\|B\bfz\|_{p}^{p}
> \lambda_{\min}(\alpha, p)n$.
\end{lemma}

\begin{lemma}\label{lem:rhobound}
Given any $\alpha$, $p$ and corresponding $\lambda_{\min}(\alpha,
p)>0$, there exists a constant $\rho^*(\alpha, p)>0$ and some
constant $c_{10}>0$ 
such that with probability at least $1-e^{-c_{10}n}$, for every
$\bfz \in \mathcal{S}$ and for every set $T \subset \{1,2,...,m\}$
with $|T| \leq \rho^*(\alpha, p) m$, $\|B_T\bfz\|_p^{p} <
\frac{1}{2}\lambda_{\min}(\alpha, p)n$.
\end{lemma}

We defer the proofs of Lemma \ref{lem:lambdamin} and Lemma
\ref{lem:rhobound} for later discussion, and first present our
result on bounds for strong recovery of $\ell_p$-minimization with
given $\alpha \in (0,1)$.

\begin{theorem}\label{thm:boundalpha}
For any $0<p \leq 1$, for matrix $A^{m \times n}$
($\alpha=\frac{m}{n}$) with i.i.d $\mathcal{N}(0,1)$ entries, there
exists a constant $c_{11}>0$ 
such that with probability at least $1-e^{-c_{11}n}$, $\bfx$ is the
unique solution to the $\ell_p$-minimization problem (\ref{eqn:lp})
for every vector $\bfx$ up to $\rho^*(\alpha, p)n$-sparse.
\end{theorem}

\begin{proof}
Let $\mathcal{S}$ be the unit sphere in $\mathcal{R}^{n-m}$. Then
\begin{eqnarray}
&& P(\textrm{Strong recovery succeeds to recover vectors
up to } \rho^*(\alpha, p) n \textrm{-sparse}) \nonumber\\
&=&P(\forall \textrm{ non-zero }\bfz \in \mathcal{R}^{n-m}, \forall T \textrm{ with } |T| =\rho^*(\alpha, p)n,  \|B_{T}\bfz\|_p^{p} < \frac{1}{2}\|B\bfz\|_p^{p}) \nonumber\\
&=&P(\forall \bfz \in \mathcal{S}, \forall T \textrm{ with } |T| =\rho^*(\alpha, p)n,  \|B_{T}\bfz\|_p^{p} < \frac{1}{2}\|B\bfz\|_p^{p}) \nonumber\\
& \geq & P(\forall \bfz \in \mathcal{S}, \forall T \textrm{ with } |T| =\rho^*(\alpha, p)n, \|B_{T}\bfz\|_p^{p} < \frac{1}{2}\lambda_{\min}(\alpha, p)n ,\textrm{ and }\|B\bfz\|_p^{p} > \lambda_{\min}(\alpha, p)n)  \nonumber\\
& \geq & 1- P(\exists \bfz \in \mathcal{S}, \textrm{ s.t. }
\|B\bfz\|_p^{p} \leq \lambda_{\min}(\alpha, p)n) \nonumber\\
&&- P(\exists \bfz \in \mathcal{S},  \exists T \textrm{ with } |T|=
\rho^*(\alpha,p)n \textrm{ s.t. }
\|B_T\bfz\|_p^{p} \geq \lambda_{\min}(\alpha, p)n/2) \nonumber\\
& =&1- e^{-c_{9}n}-e^{-c_{10}n}, \label{eqn:c910}
\end{eqnarray}
where the first equality follows from Theorem \ref{thm:slp}, the
second equality holds since for any non-zero $\bfz \in
\mathcal{R}^{n-m}$, $\bfz/\|\bfz\|_2 \in \mathcal{S}$. From  Lemma
\ref{lem:lambdamin} we know there exists $c_{9}>0$ such that
$P(\exists \bfz \in \mathcal{S}, \textrm{ s.t. } \|B\bfz\|_p^{p}
\leq \lambda_{\min}(\alpha, p)n) \leq e^{-c_9n}$, and from Lemma
\ref{lem:rhobound} we know there exists $c_{10}>0$ such that
$P(\exists \bfz \in \mathcal{S},  \exists T \textrm{ s.t. }
\|B_T\bfz\|_p^{p} \geq \frac{1}{2}\lambda_{\min}(\alpha, p)n )\leq
e^{-c_{10}n}$, then there exists $c_{11}>0$ which depends on
$\alpha$, $p$ and $\lambda_{\min}$ such that (\ref{eqn:c910}) $\geq
1-e^{-c_{11}n}$. Therefore, $\ell_p$-minimization can recover all
the $\rho^*(\alpha,p)n$-sparse vectors with probability at least $1-
e^{-c_{11}n}$.

\end{proof}

Theorems \ref{thm:boundalpha} implies that for every $\alpha \in
(0,1)$ and every $p \in (0,1]$, there exists a positive constant
$\rho^*(\alpha, p)$ such that $\ell_p$-minimization can recover all
the $\rho^* n$-sparse vectors with overwhelming probability. Since
$\rho^*(\alpha, p)$ is a lower bound of the threshold of the strong
recovery, we want it to be as high as possible. Next we show how to
calculate $\rho^*(\alpha, p)$ and improve it as much as possible. In
order to calculate $\rho^*(\alpha, p)$, we first calculate
$\lambda_{\min}(\alpha, p)$ in Lemma \ref{lem:lambdamin}, and then
with the obtained $\lambda_{\min}(\alpha, p)$, we can calculate
$\rho^*(\alpha, p)$ in Lemma \ref{lem:rhobound}.  We want to obtain
$\lambda_{\min}(\alpha, p)$ which is as large as possible while
Lemma \ref{lem:lambdamin} still holds, and given
$\lambda_{\min}(\alpha, p)$, we want $\rho^*(\alpha, p)$ to be as
large as possible while Lemma \ref{lem:rhobound} still holds. How to
calculate $\lambda_{\min}(\alpha, p)$ and $\rho^*(\alpha, p)$ is
stated in the following text, and Lemma \ref{lem:lambdamin} and
Lemma \ref{lem:rhobound} are proved in the meantime. The values of
$\lambda_{\min}(\alpha, p)$ and $\rho^*(\alpha, p)$ can be computed
from (\ref{eqn:lambdamin}) and (\ref{eqn:rhoalphap}).

\subsubsection{Calculation of $\lambda_{\min}(\alpha, p)$ in Lemma
\ref{lem:lambdamin}}\label{sec:lambda}\

Given $\alpha$ and $p$, define
\begin{equation}\nonumber
c_{\max}= \frac{1}{n}\sup_{ \bfz \in \mathcal{S}} \|B\bfz\|_p^p =
\frac{1}{n}\max_{ \bfz \in \mathcal{S}} \|B\bfz\|_p^p,
\end{equation}
where the second equality holds by compactness. Thus, for any
non-zero vector $\bfz$, $\|B\bfz\|_p^p \leq \|\bfz\|_p^p c_{\max}n$.
 Define
\begin{equation}\nonumber
c_{\min}
= \frac{1}{n}\min_{ \bfz \in \mathcal{S}} \|B\bfz\|_p^p.
\end{equation}
Pick a $\gamma$-net $\Sigma_2$ of $\mathcal{S}$ with cardinality at
most $(1+2/\gamma)^{n-m}$ \cite{Ledoux01} and $\gamma>0$ to be
chosen later, we define
\begin{equation}\nonumber
\theta= \frac{1}{n}\min_{ \bfz \in \Sigma_2} \|B\bfz\|_p^p.
\end{equation}
Then for every $\bfz \in \mathcal{S}$, there exists $\bfz' \in
\Sigma_2$ such that $\|\bfz-\bfz'\|_2 \leq \gamma$. We have
\begin{eqnarray}\label{eqn:bz}
\|B\bfz\|_p^p \geq  \|B\bfz'\|_p^p -\|B(\bfz-\bfz')\|_p^p 
 \geq  \theta n - \gamma^p c_{\max} n,
\end{eqnarray}
where the first inequality follows from triangular inequality and
the second inequality follows from the definition of $c_{\max}$.
Since (\ref{eqn:bz}) holds for every $\bfz$ in $\mathcal{S}$, we
have
\begin{equation} \label{eqn:netmin}
c_{\min} \geq \theta-\gamma^pc_{\max}.
\end{equation}

To calculate  $\lambda_{\min}(\alpha, p)$, we essentially need to
characterize $c_{\min}$. From (\ref{eqn:netmin}), we can achieve
this by characterizing $\theta$ and $c_{\max}$.

We first show that there exists constant $b>0$ such that with
overwhelming probability, $\theta > b$ holds, i.e. $\|B\bfz\|_p^p
> bn$ for all $\bfz$ in $\Sigma_2$. 
\begin{eqnarray}
\hspace{-0.18in}&&P(\theta \leq b)= P(\exists \bfz \in \Sigma_2
\textrm{ s.t. }
\|B\bfz\|_p^p \leq bn) \nonumber\\
\hspace{-0.18in}& \leq & \sum_{ \bfz \in \Sigma_2}P(\|B \bfz\|_p^p \leq bn) \nonumber\\
\hspace{-0.18in}& \leq & (1+2/\gamma)^{n-m} e^{tbn}E[e^{-t\sum_i
|B_i\bfz|^p}], \quad  \forall t>0\nonumber\\
\hspace{-0.18in}& =& (1+2/\gamma)^{(1-\alpha)n} e^{tbn}E[e^{-t|X|^p}]^n, \quad \forall t>0\nonumber\\
\hspace{-0.18in}&=&
e^{\large((1-\alpha)\log(1+2/\gamma)+\log(E[e^{-t|X|^p}])+bt\large)n},
\quad \forall t>0, \label{eqn:theta}
\end{eqnarray}
where $X \sim \mathcal{N}(0,1)$. The first inequality follows from
the union bound and the fact that $P(\|B \bfz\|_p^p \leq bn)$ is the
same for all $\bfz \in \Sigma_2$ since $B$ has i.i.d.
$\mathcal{N}(0,1)$ entries. The second inequality follows from the
Chernoff bound. Note that
\begin{eqnarray}
E[e^{-t|X|^p}]&=&\sqrt{2/\pi}\int_0^{\infty}e^{-tx^p}e^{-\frac{1}{2}x^2}dx
\nonumber\\
&=&t^{-\frac{1}{p}}\sqrt{2/\pi}\int_0^{\infty}e^{-y^p}e^{-\frac{1}{2}(t^{-\frac{1}{p}}y)^2}dy.\label{eqn:expt}\\
& \leq & t^{-\frac{1}{p}}\sqrt{2/\pi}\int_0^{\infty}e^{-y^p}dy
\nonumber\\
&=&t^{-\frac{1}{p}}\sqrt{2/\pi}\Gamma(1/p)/p, \label{eqn:expt1}
\end{eqnarray}
where (\ref{eqn:expt}) holds from changing variables using
$x=t^{-\frac{1}{p}}y$, and the inequality follows from the fact that
$e^{-\frac{1}{2}(t^{-\frac{1}{p}}y)^2}\leq 1$ for all $y\geq 0$. If
it further holds that $t>1$, then $t^{-\frac{1}{p}}<1$. Then from
(\ref{eqn:expt}) we have
\begin{equation}\label{eqn:tbound}\nonumber
E[e^{-t|X|^p}] \geq t^{-\frac{1}{p}}\sqrt{2/\pi}
\int_0^{\infty}e^{-y^p-\frac{1}{2}y^2}dy.
\end{equation}
Since $\int_0^{\infty}e^{-y^p-\frac{1}{2}y^2}dy$ exists and is
positive, then combining (\ref{eqn:expt1}) and (\ref{eqn:tbound}),
we have
\begin{equation}\label{eqn:bigo}
E[e^{-t|X|^p}]=O(t^{-\frac{1}{p}}).
\end{equation}
Since (\ref{eqn:theta}) holds for all $t>0$, 
we let
$t=\gamma^{-p(1-\alpha+\epsilon)}$ for any $\epsilon$ such that
$0<\epsilon \leq \alpha$ and let $b(\gamma)=1/t$,
%
then from
(\ref{eqn:theta}) we have
\begin{equation}\nonumber 
P(\theta \leq b(\gamma)) \leq
e^{\large((1-\alpha)\log(1+2/\gamma)+\log(O(\gamma^{1-\alpha+\epsilon}))+1\large)n}=e^{-\kappa
n},
\end{equation}
where $\kappa(\gamma)=-(1-\alpha)\log(1+\frac{2}{\gamma})-\log(O(\gamma^{1-\alpha+\epsilon}))-1$. Note that since $\epsilon>0$, 
when $\gamma$ is sufficiently small, $\kappa(\gamma)>0$. Therefore
when $\gamma \leq \xi$ for some small $\xi>0$, there exists constant
$\kappa(\gamma)>0$ such that
\begin{equation}\label{eqn:thetab}
P(\theta \leq b(\gamma)=\gamma^{p(1-\alpha+\epsilon)})\leq
e^{-\kappa(\gamma) n}.
\end{equation}

We next show that there exists some $\lambda_{\max}(\alpha,p)>0$
such that with overwhelming probability,
$c_{\max}<\lambda_{\max}(\alpha, p)$ holds. In fact, we have the
following Lemma:
\begin{lemma}\label{lem:lambdamax}
Given any $\alpha$ and $p$, there exists a constant
$\lambda_{\max}(\alpha, p)>0$ and some constant $c_{12}>0$
such that with probability at least $1-e^{-c_{12}n}$, for every
$\bfz \in \mathcal{S}$, $\|B\bfz\|_{p}^{p} < \lambda_{\max}(\alpha,
p)n$.
\end{lemma}

Lemma \ref{lem:lambdamax} indicates that there exists
$\lambda_{\max}(\alpha, p)$ and $c_{12}>0$ such that
\begin{equation}\label{eqn:cmlm}
P(c_{\max} < \lambda_{\max}(\alpha, p)) \geq 1-e^{-c_{12}n}.
\end{equation}
Please refer to the Appendix for the calculation of
$\lambda_{\max}(\alpha, p)$, and Lemma \ref{lem:lambdamax} is proved
in the meantime. In order to obtain a good bound of recovery
threshold, we want $\lambda_{\max}(\alpha, p)$ to
 be as small as possible while Lemma \ref{lem:lambdamax} still
 holds. The numerical value of $\lambda_{\max}(\alpha, p)$ can be
 computed from (\ref{eqn:lambdamax}).

 Then after characterizing $\theta$ and $c_{\max}$ separately, we
 are ready to characterize $c_{\min}$.
\begin{eqnarray}
&&P( c_{\min}\leq \gamma^{p(1-\alpha+\epsilon)}-\gamma^p
\lambda_{\max}(\alpha, p)) \nonumber\\ &\leq&  P(
\theta-\gamma^pc_{\max}\leq \gamma^{p(1-\alpha+\epsilon)}-\gamma^p
\lambda_{\max}(\alpha, p))
\nonumber\\
&\leq& P(\theta \leq \gamma^{p(1-\alpha+\epsilon)})+P(c_{\max} \geq \lambda_{\max}(\alpha, p)) \nonumber\\
& \leq & e^{-\kappa n}+ e^{-c_{12}n}, \nonumber
\end{eqnarray}
where the first inequality follows from (\ref{eqn:netmin}), and the
last inequality follows from (\ref{eqn:thetab}) and
(\ref{eqn:cmlm}). Then for any $\gamma \leq \xi$, there exists
constant $c_9>0$ such that $ P(c_{\min}\leq
\gamma^{p(1-\alpha+\epsilon)}-\gamma^p
\lambda_{\max}(\alpha, p)) \leq e^{-c_9 n}$. Given $\lambda_{\max}(\alpha,p)$, 
let
\begin{equation}\label{eqn:lambdamin}
\lambda_{\min}(\alpha,p)=\max \limits_{0<\gamma \leq \xi}
\gamma^{p(1-\alpha+\epsilon)}-\gamma^p \lambda_{\max}(\alpha,p).
\end{equation}
 Note that
since $1-\alpha+\epsilon<1$, $\gamma^{p(1-\alpha+\epsilon)}-\gamma^p
\lambda_{\max}>0$ when $\gamma$ is sufficiently small, therefore
$\lambda_{\min}>0$, and Lemma \ref{lem:lambdamin} follows.
%
%
%

\subsubsection{Calculation of $\rho^*(\alpha, p)$ in Lemma
\ref{lem:rhobound}}  \

 For any given set $T \subset
\{1,2,...,n\}$ with $|T|=\rho n$
($0<\rho<1$), 
define
\begin{equation}\nonumber
d_{\max}
= \frac{1}{n}\max_{ \bfz \in \mathcal{S}} \|B_T\bfz\|_p^p.
\end{equation}
 Given a $\gamma$-net $\Sigma_3$ of $\mathcal{S}$ with
cardinality at most $(1+2/\gamma)^{n-m}$ and $\gamma>0$ to be chosen
later, define
\begin{equation}\nonumber
\tau = \frac{1}{n}\max_{ \bfz \in \Sigma_3} \|B_T\bfz\|_p^p.
\end{equation}
Then for every $\bfz \in \mathcal{S}$, there exists $\bfz' \in
\Sigma_3$ such that $\|\bfz-\bfz'\|_2 \leq \gamma$. Then for every
$\bfz \in \mathcal{S}$, we have
$\|B_T\bfz\|_p^p \leq  \|B_T\bfz'\|_p^p +\|B_T(\bfz-\bfz')\|_p^p 
 \leq  \tau n + \gamma^p d_{\max} n$. 
Thus,
\begin{equation}\label{eqn:dtau}
d_{\max} \leq \tau/(1-\gamma^p).
\end{equation}

Given $\lambda_{\min}(\alpha, p)$ (denoted by $\lambda_{\min}$ here
for simplicity), in order to obtain $\rho^*(\alpha, p)$ such that
Lemma \ref{lem:rhobound} holds, we essentially need to find $\rho$
such that for any $T$ with its corresponding $d_{\max}$, with
overwhelming probability $d_{\max}<\lambda_{\min}/2$ holds for all
$T$ with $|T|=\rho m$ at the same time. From
(\ref{eqn:dtau}), we first consider the probability that $\tau \geq \lambda_{\min}(1-\gamma^p)/2$ holds for a given set $T$. 
\begin{eqnarray}
&&P(\tau \geq \lambda_{\min}(1-\gamma^p)/2, \textrm{ given }
T)\nonumber\\&= &P(\exists \bfz \in \Sigma_3\textrm{ s.t. }
\|B_T\bfz\|_p^p \geq \lambda_{\min}(1-\gamma^p)n/2) \nonumber\\
& \leq
& \sum_{ \bfz \in \Sigma_3} P(\|B_T\bfz\|_p^p \geq \frac{\lambda_{\min}(1-\gamma^p)n}{2}) \nonumber \\
&=& \sum_{ \bfz \in \Sigma_3}P(\sum_{i\in T} |B_i \bfz|^p \geq \frac{\lambda_{\min}(1-\gamma^p)n}{2}) \nonumber\\
& \leq & (1+2/\gamma)^{n-m} \min_{t>0}
e^{-t\lambda_{\min}(1-\gamma^p)n/2} E[e^{t\sum_{i \in T}
|B_i\bfz|^p}]\nonumber\\
& =& (1+2/\gamma)^{(1-\alpha)n}\min_{t>0} e^{-t\lambda_{\min}(1-\gamma^p)n/2}E[e^{t|X|^p}]^{\rho n}\nonumber\\
&=& e^{\large((1-\alpha)\log(1+\frac{2}{\gamma})+\min
\limits_{t>0}(\rho\log(E[e^{t|X|^p}])-t\lambda_{\min}(1-\gamma^p)/2)\large)n},
\label{eqn:tau}
\end{eqnarray}
where $X \sim \mathcal{N}(0,1)$, the first inequality follows from
the union bound and the fact that  the second inequality follows
from the Chernoff bound. Note that since $B$ has i.i.d.
$\mathcal{N}(0,1)$ entries, (\ref{eqn:tau}) holds for any $T$ as
long as $|T| = \rho n$.

Given $\rho$, $\lambda_{\min}$ and $\gamma$, since the second
derivative of
$\rho\log(E[e^{t|X|^p}])-t\lambda_{\min}(1-\gamma^p)/2$ to $t$ is
positive, then its minimum is achieved where its first derivative is
0.
\begin{eqnarray}
0&=&\frac{d[\rho\log(E[e^{t|X|^p}])-t\lambda_{\min}(1-\gamma^p)/2]}{dt} \nonumber \\
&=&\frac{d}{dt}(\rho\log(\sqrt{\frac{2}{\pi}}\int_0^{\infty}e^{tx^p-\frac{1}{2}x^2}dx)-t\lambda_{\min}(1-\gamma^p)/2) \nonumber \\
&=& \frac{\rho\int_0^{\infty}x^p e^{tx^p-\frac{1}{2}x^2} dx}{
\int_0^{\infty}e^{tx^p-\frac{1}{2}x^2
}dx}-\lambda_{\min}(1-\gamma^p)/2 \label{eqn:minimum2}.
\end{eqnarray}
Note that when $\rho<\lambda_{\min}(1-\gamma^p)/(2E[|X|^p])$, the
solution of $t$ to (\ref{eqn:minimum2}) is always
positive, thus it is also the solution to $\min_{t>0}(\rho\log(E[e^{t|X|^p}])-t\lambda_{\min}(1-\gamma^p)/2)$. 
Now consider the probability that $\|B_T\bfz\|^{p} \geq
\frac{1}{2}\lambda_{\min}n$ for some $\bfz \in \mathcal{S}$ and $T$
with $|T| = \rho n$.
\begin{eqnarray}
&&P( \exists \bfz \in \mathcal{S}, \exists T \textrm{ s.t. } |T| = \rho n, \|B_T \bfz \|_p^{p} \geq \lambda_{\min}n/2) \nonumber \\
&\leq & {{n}\choose{\rho n}} P( \exists \bfz \in \mathcal{S}
\textrm{ s.t. } \|B_T\bfz\|_p^{p} \geq \lambda_{\min}n/2,
\nonumber\\
&& \textrm{ for given } T \subset \{1,2,...,n\} \textrm{ and }
|T|=\rho
n) \nonumber\\
&= & {{n}\choose{\rho n}} P( d_{\max} \geq \lambda_{\min}/2) \nonumber \\
& \leq &{{n}\choose{\rho n}} P( \tau/(1-\gamma^p) \geq \lambda_{\min}/2) \nonumber \\
&=& {{n}\choose{\rho n}} P( \tau \geq \lambda_{\min}(1-\gamma^p)/2) \nonumber \\
&\leq & 2^{nH(\rho)}e^{\Large((1-\alpha)\log(1+2/\gamma)+\min\limits_{t>0}(\rho\log(E[e^{t|X|^p}])-t\lambda_{\min}(1-\gamma^p)/2)\Large)n} \nonumber\\
&=& e^{ \Large(H(\rho)\log
2+(1-\alpha)\log(1+2/\gamma)+\min\limits_{t>0}(\rho\log(E[e^{t|X|^p}])-t\lambda_{\min}(1-\gamma^p)/2)\Large)n},
\label{eqn:pfbound}
\end{eqnarray}
where the first inequality follows from the union bound and the
second inequality follows from (\ref{eqn:dtau}). Note that given
$\alpha$, $p$, and $\lambda_{\min}$, for every $\gamma$, as $\rho
\rightarrow 0$, $H(\rho)$ goes to 0, and $\min
\limits_{t>0}(\rho\log(E[e^{t|X|^p}])-t\lambda_{\min}(1-\gamma^p)/2$
goes to $-\infty$, thus, there exists $\rho(\alpha, p, \gamma)>0$
such that the exponent of (\ref{eqn:pfbound}) is negative for all
$\rho \leq\rho(\alpha, p, \gamma)$. In other words, for each
$\gamma$, there exists some $c_{10}>0$ such that (\ref{eqn:pfbound})
$\leq e^{-c_{10}n}$ when $\rho=\rho(\alpha, p, \gamma)$.
Then, with probability at least $1-e^{-c_{10}n}$, for every $\bfz
\in \mathcal{S}$ and for every set $T \subset \{1,2,...,n\}$ with
$|T|
\leq \rho(\gamma) n$, $\|B_T\bfz\|_p^{p} < \lambda_{\min}n/2$. 
Let
\begin{equation}\label{eqn:rhoalphap}
\rho^*(\alpha, p)=\max_{\gamma} \rho(\alpha, p, \gamma),
\end{equation}
then Lemma \ref{lem:rhobound} follows.

\begin{figure}[t]
      \centering
      \includegraphics[scale=0.5]{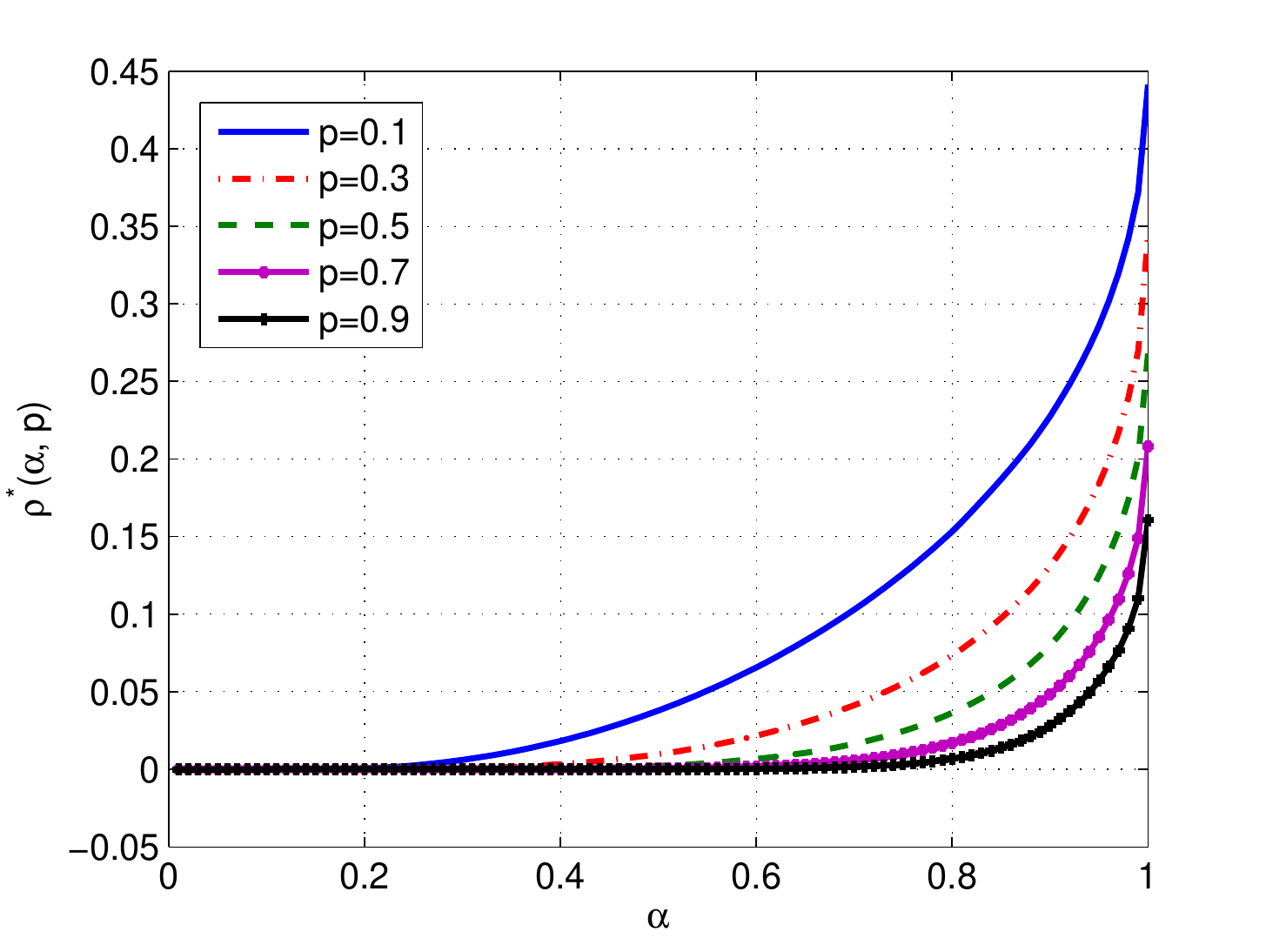}
 \caption{$\rho^*(\alpha,p)$ against $\alpha$ for different $p$}
   \label{fig:p}
\end{figure}
\begin{figure}[t]
      \centering
      \includegraphics[scale=0.5]{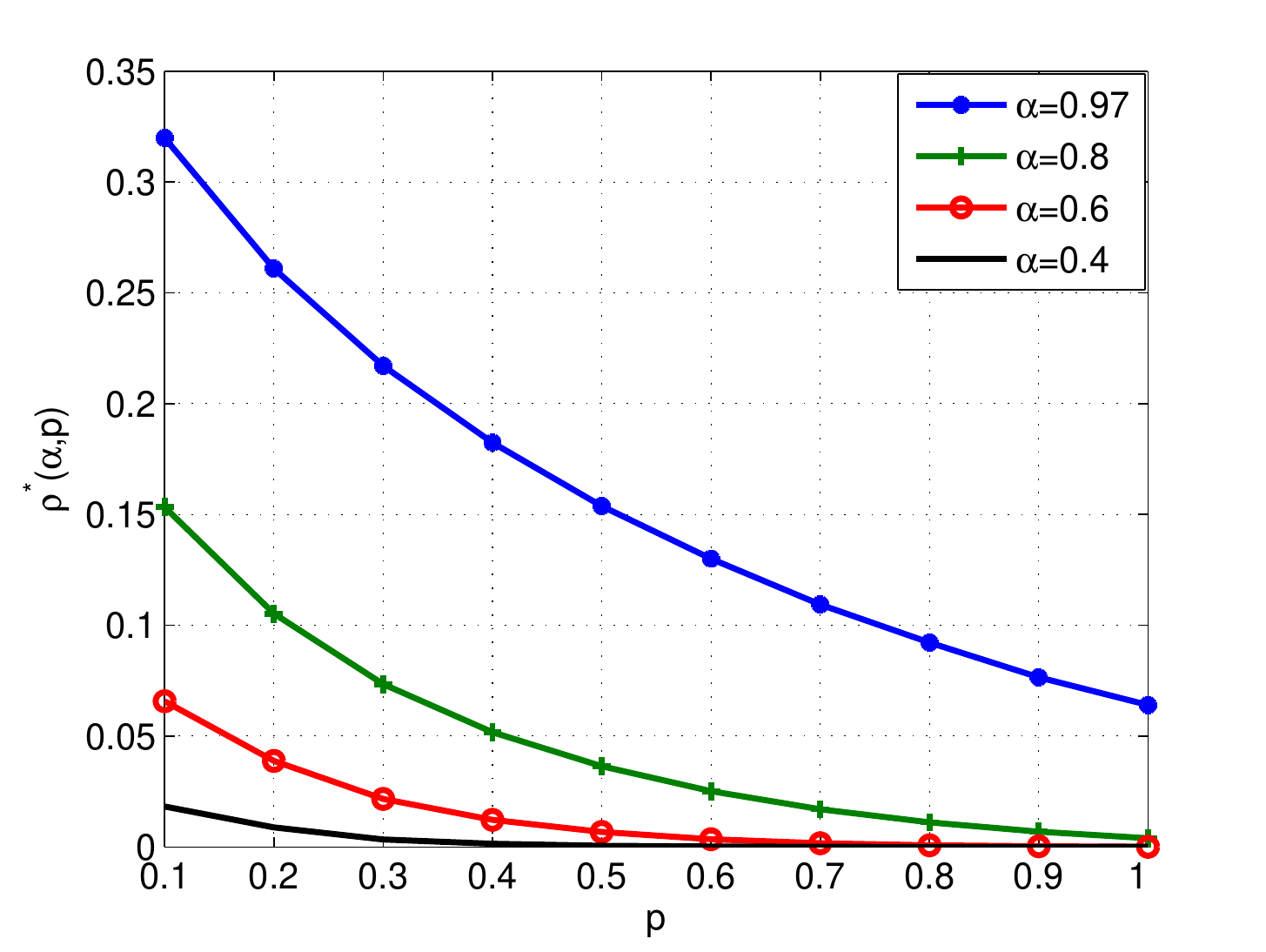}
 \caption{$\rho^*(\alpha,p)$ against $p$ for different $\alpha$}
   \label{fig:alpha}
\end{figure}

Theorem \ref{thm:boundalpha} establishes the existence of
$\rho^*(\alpha,p)>0$ for all $0<\alpha<1$ and $0<p\leq 1$ such that
$\ell_p$-minimization can recover all the $\rho^*(\alpha,
p)n$-sparse vectors with overwhelming probability. We numerically
calculate this bound by calculating first $\lambda_{\max}(\alpha,
p)$ in Lemma \ref{lem:lambdamax} from (\ref{eqn:lambdamax}), and
then $\lambda_{\min}(\alpha, p)$ in Lemma \ref{lem:lambdamin} from
(\ref{eqn:lambdamin}), and finally $\rho^*(\alpha, p)$ in Lemma
\ref{lem:rhobound} from (\ref{eqn:rhoalphap}). Fig. \ref{fig:p}
shows the curve of $\rho^*(\alpha,p)$ against $\alpha$ for different
$p$, and Fig. \ref{fig:alpha} shows the curve of $\rho^*(\alpha,p)$
against $p$ for different $\alpha$. Note that for any $p$,
$\lim_{\alpha \rightarrow 1} \rho^*(\alpha,p)$ is slightly smaller
than the limiting threshold of strong recovery we obtained in
Section \ref{sec:sbd}. For example, when $p=0.5$, the threshold
$\rho^*(0.5)$ we obtained in Section \ref{sec:sbd} is 0.3406, and
the bound $\rho^*(\alpha, 0.5)$ we obtained here is approximately
0.268 when $\alpha$ goes to 1. This is because in Section
\ref{sec:sbd} we employed a finer technique to characterize the sum
of the largest $\rho n$ terms of $n$ i.i.d. random variables
directly, while in Section \ref{sec:sbdfinite} introducing the union
bound causes some slackness.

Compared with the bound obtained in \cite{BCT09} through restricted
isometry condition, our bound $\rho^*(\alpha, p)$ is tighter when
$\alpha$ is relatively large. For example, when $p=1$, the bound in
\cite{BCT09} (Fig.3.2(a)) is in the order of $10^{-3}$ for all
$\alpha\in(0,1)$ and upper bounded by $0.0035$, while
$\rho^*(\alpha, 1)$ is greater than $0.0039$ for all $\alpha\geq
0.8$ and increases to 0.1308 as $\alpha \rightarrow 1$. When
$p=0.5$, the bound in \cite{BCT09} (Fig.3.2(c)) is in the order of
$10^{-3}$ for all $\alpha\in(0,1)$ and upper bounded by $0.01$,
while here $\rho^*(\alpha, 0.5)$ is greater than $0.011$ for all
$\alpha\geq 0.65$ and increases to 0.268 as $\alpha \rightarrow 1$.
Therefore, although \cite{BCT09} provides a better bound than ours
when $\alpha$ is small, our bound $\rho^*$ improves over that in
\cite{BCT09} when $\alpha$ is relatively large. \cite{Donoho06}
applies geometric face counting technique to the strong bound of
successful recovery of $\ell_1$-minimization (Fig.1.1). Since if the
necessary and sufficient condition (\ref{eqn:slp}) is satisfied for
$p=1$, then it is also satisfied for all $p<1$, therefore the bound
in \cite{DoT09} can serve as the bound of successful recovery for
all $0<p<1$. Our bound $\rho^*(\alpha, p)$ in Section
\ref{sec:finite} is higher than that in \cite{Donoho06} when
$\alpha$ is relatively large.

\subsection{Weak Recovery}\label{sec:wbdfinite}
Theorem \ref{thm:wb} provides a sufficient condition for successful
recovery of every $\rho n$-sparse vector $\bfx$ on one support $T$
with one sign pattern, which requires $\|B_{T^
-}\bfz\|_p^p<\|B_{T^c}\bfz\|_p^p$ to hold for all non-zero $\bfz \in
\mathcal{R}^n$, where given $\bfz$, $T^-=\{ i : B_i \bfz x_i <0\}$.
Given $\alpha$, $p$ and $\rho \in (0,1)$, we will establish a lower
bound of $\|B_{T^c}\bfz\|_p^p$ for all $\bfz \in \mathcal{S}$ in
Lemma \ref{lem:lambdaminw}, and establish an upper bound of
$\|B_{T^-}\bfz\|_p^p$ in Lemma \ref{lem:lambdamaxw}. If there exists
$\rho_w^*(\alpha,p)>0$ such that the corresponding lower bound of
$\|B_{T^c}\bfz\|_p^p$ is greater than the upper bound of
$\|B_{T^-}\bfz\|_p^p$, which in fact is always true as we will see
in Theorem \ref{thm:wboundalpha}, then $\rho^*_w(\alpha,p)$ serves
as a lower bound of recovery threshold of $\ell_p$-minimization for
vectors on a fixed support with a fixed sign pattern.

The technique to establish the lower bound of $\|B_{T^c}\bfz\|_p^p$
for all $\bfz \in \mathcal{S}$ is the same as that in Lemma
\ref{lem:lambdamin}. We state the result in Lemma
\ref{lem:lambdaminw}, please refer to the appendix for its proof.
\begin{lemma}\label{lem:lambdaminw}
Given $\alpha$, $p$ and set $T \subset \{1,...,n\}$ with $|T|=\rho
n$, with probability at least $1-e^{-c_{13}n}$ for some $c_{13}>0$,
for all $\bfz \in \mathcal{S}$, $\|B_{T^c}\bfz\|_p^p
<(1-\rho)\lambda_{\max}(\frac{\alpha-\rho}{1-\rho},p)n$,
and with probability at least $1-e^{-c_{14}n}$ for some $c_{14}>0$,
for all $\bfz \in \mathcal{S}$, $\|B_{T^c}\bfz\|_p^p
>(1-\rho)\lambda_{\min}(\frac{\alpha-\rho}{1-\rho},p)n$,
where $\lambda_{\max}(\alpha,p)$ and $\lambda_{\min}(\alpha,p)$ are
defined in (\ref{eqn:lambdamax}) and (\ref{eqn:lambdamin})
respectively.
\end{lemma}

Given $T$ with $|T| = \rho n$, Lemma \ref{lem:lambdaminw} provides a
lower bound of $\|B_{T^c}\bfz\|_p^p$ which holds with overwhelming
probability for
all $\bfz \in \mathcal{S}$. Please refer to the Appendix for its proof. 
Next we will provide an upper bound of $\|B_{T^-}\bfz\|_p^p$ for all
$\bfz \in \mathcal{S}$ in Lemma \ref{lem:lambdamaxw}. One should be
cautious that the set $T^-$ varies for different $\bfz$. To improve
the bound of the threshold of successful weak recovery, we want
$\tilde{\lambda}_{\max}(\alpha,p,\rho)$ to be as small as possible
while Lemma \ref{lem:lambdamaxw} still holds.
$\tilde{\lambda}_{\max}(\alpha,p,\rho)$ can be computed from
(\ref{eqn:tildelambdamax}), please refer to the Appendix for its
detailed calculation.

\begin{lemma}\label{lem:lambdamaxw}
Given $\alpha$, $p$ and set $T \subset \{1,...,n\}$ with $|T|=\rho
n$, with probability at least $1-e^{-c_{15}n}$ for some $c_{15}>0$,
for every $\bfz \in \mathcal{S}$, $\|B_{T^-}\bfz\|_p^p <\rho
\tilde{\lambda}_{\max}(\alpha,p,\rho)n$, for some
$\tilde{\lambda}_{\max}(\alpha,p,\rho)>0$.
\end{lemma}

With the help of Lemma \ref{lem:lambdaminw} and Lemma
\ref{lem:lambdamaxw}, we are ready to present the result regarding
the lower bound of recovery threshold via $\ell_p$-minimization in
the weak sense for given $\alpha$.

\begin{theorem}\label{thm:wboundalpha}
For any $0<p \leq 1$, for matrix $A^{m \times n}$ with i.i.d
$\mathcal{N}(0,1)$ entries, there exists constant $\rho_w^*(\alpha,
p)>0$ and $c_{16}>0$ such that with probability at least
$1-e^{-c_{16}n}$, $\bfx$ is the unique solution to the
$\ell_p$-minimization problem (\ref{eqn:lp}) for every
$\rho_w^*(\alpha, p)n$-sparse vector $\bfx$ on one support $T$ with
one sign pattern.
\end{theorem}

\begin{proof}
Note that given $p$ and $\alpha$, since
$\tilde{\lambda}_{\max}(\alpha,p,\rho)$ and
$\lambda_{\min}(\frac{\alpha-\rho}{1-\rho},p)$ are both positive for
all $\rho\in (0,1)$, and one can check from the definition of
$\tilde{\lambda}_{\max}(\alpha,p,\rho)$ and
$\lambda_{\min}(\frac{\alpha-\rho}{1-\rho},p)$ that when $\rho$
decreases, $\tilde{\lambda}_{\max}(\alpha,p,\rho)$ is
non-increasing, and $\lambda_{\min}(\frac{\alpha-\rho}{1-\rho},p)$
is non-decreasing. Therefore, there always exists $\rho_w^*(\alpha,
p)>0$ (denoted by $\rho_w^*$ for simplicity here) such that
\begin{equation}\label{eqn:rhow'}
\rho_w^*\tilde{\lambda}_{\max}(\alpha,p,\rho_w^*) \leq (1-\rho_w^*)
\lambda_{\min}(\frac{\alpha-\rho_w^*}{1-\rho_w^*},p).
\end{equation}
Now consider the probability that $\ell_p$-minimization can recover
all the $\rho_w^* n$-sparse $\bfx$ on one fixed support $T$ with one
fixed sign pattern. From Theorem \ref{thm:wb} we know that
$\|B_{T^-}\bfz\|_p^{p} < \|B_{T^c}\bfz\|_p^{p}$ for all non-zero
$\bfz\in \mathcal{R}^{n-m}$ is a sufficient condition for the
success of weak recovery, thus
\begin{eqnarray}
&& P(\textrm{Weak recovery succeeds
up to } \rho_w^* n \textrm{-sparse}) \nonumber\\
&\geq &P(\forall \textrm{ non-zero }\bfz \in \mathcal{R}^{n-m},  \|B_{T^-}\bfz\|_p^{p} < \|B_{T^c}\bfz\|_p^{p}) \nonumber\\
&=&P(\forall \bfz \in \mathcal{S}, \|B_{T^-}\bfz\|_p^{p} < \|B_{T^c}\bfz\|_p^{p}) \nonumber\\
& \geq & P(\forall \bfz \in \mathcal{S},
\|B_{T^-}\bfz\|_p^{p} < \rho_w^*\tilde{\lambda}_{\max}(\alpha,p,\rho_w^*),\textrm{ and }   \nonumber\\
&& \|B_{T^c}\bfz\|_p^{p} > (1-\rho_w^*)
\lambda_{\min}(\frac{\alpha-\rho_w^*}{1-\rho_w^*},p)) \nonumber\\
& \geq &1- e^{-c_{15}n}-e^{-c_{14}n}, \label{eqn:c1415}
\end{eqnarray}
where the equality holds since for any non-zero $\bfz \in
\mathcal{R}^{n-m}$, $\bfz/\|\bfz\|_2 \in \mathcal{S}$, and the
second inequality follows from (\ref{eqn:rhow'}). From Lemma
\ref{lem:lambdaminw} we know there exists  $c_{14}>0$ such that
$P(\|B_{T^c}\bfz\|_p^{p} > (1-\rho_w^*)
\lambda_{\min}(1-\frac{1-\alpha}{1-\rho_w^*},p)) \geq 1-
e^{-c_{14}n}$, and from Lemma \ref{lem:lambdamaxw} we know there
exists $c_{15}>0$ such that $P(\forall \bfz \in \mathcal{S},
\|B_{T^-}\bfz\|_p^{p} <
\rho_w^*\tilde{\lambda}_{\max}(\alpha,p,\rho_w^*)) \geq 1-
e^{-c_{14}n}$, then (\ref{eqn:c1415}) holds.
%
%
Thus, there exists $c_{16}>0$ such that with probability at least
$1-e^{-c_{16}n}$, $\ell_p$-minimization problem can recover all
$\rho^*_w n$-sparse vectors on fixed support $T$ with fixed sign
pattern.
\end{proof}

\begin{figure}[t]
      \centering
      \includegraphics[scale=0.5]{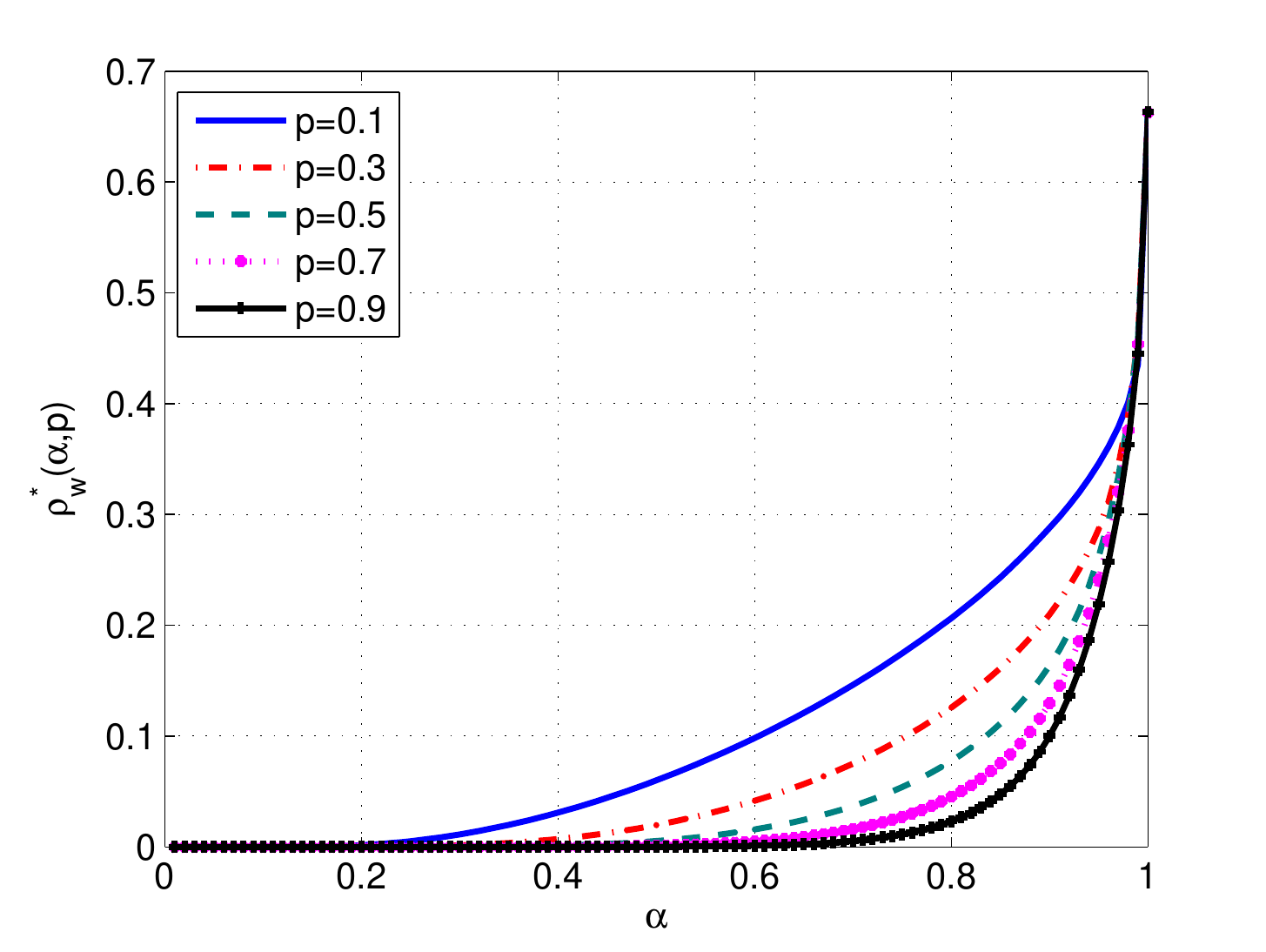}
 \caption{$\rho^*_w(\alpha,p)$ against $\alpha$ for different $p$}
   \label{fig:pw}
\end{figure}

\begin{figure}[t]
      \centering
      \includegraphics[scale=0.5]{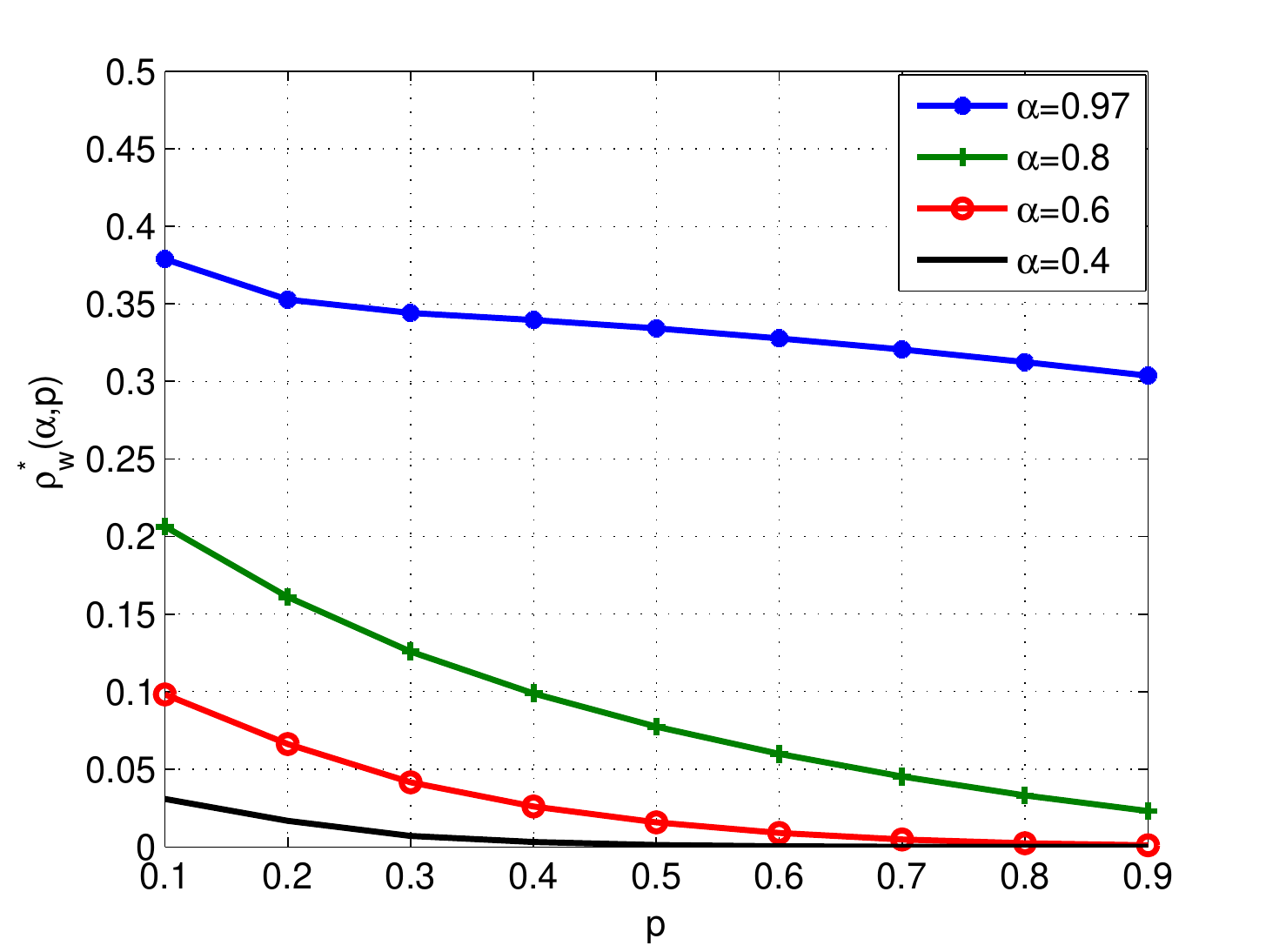}
 \caption{$\rho^*_w(\alpha,p)$ against $p$ for different $\alpha$}
   \label{fig:alphaw}
\end{figure}

Theorem \ref{thm:wboundalpha} establishes the existence of a
positive bound $\rho^*_w(\alpha,p)$ and defines $\rho^*_w(\alpha,p)$
in (\ref{eqn:rhow'}). To obtain $\rho^*_w(\alpha,p)$, we first
calculate $\lambda_{\min}(\frac{\alpha-\rho}{1-\rho},p)$ in Lemma
\ref{lem:lambdaminw} from (\ref{eqn:lambdamin}) and
$\tilde{\lambda}_{\max}(\alpha, p, \rho)$ in Lemma
\ref{lem:lambdamaxw} from (\ref{eqn:tildelambdamax}) for every
$\rho$, then find the largest $\rho^*_w(\alpha,p)$ such that
(\ref{eqn:rhow'}) holds. We numerically calculate this bound and
illustrate the results in Fig. \ref{fig:pw} and Fig.
\ref{fig:alphaw}. Fig. \ref{fig:pw} shows the curve of
$\rho^*_w(\alpha, p)$ against $\alpha$ for different $p$, and Fig.
\ref{fig:alphaw} shows the curve of $\rho^*_w(\alpha, p)$ against
$p$ for different $\alpha$. When $\alpha \rightarrow 1$,
$\rho^*_w(\alpha, p)$ goes to $2/3$ for all $p \in (0,1)$, which
coincides with the limiting threshold discussed in Section
\ref{sec:wbd}. As indicated in Fig. 1.2 of \cite{Don06}, the weak
recovery threshold of $\ell_1$-minimization is greater than 2/3 for
all $\alpha$ that is greater than 0.9, since the weak recovery
threshold of $\ell_p$-minimization ($p \in [0,1)$) when $\alpha
\rightarrow 1$ is all 2/3, therefore for all $\alpha
>0.9$, the weak recovery threshold of $\ell_1$-minimization is
greater than that of $\ell_p$-minimization for all $p \in [0,1)$.

\section{$\ell_1$-minimization can perform better than $\ell_p$-minimization ($p \in
[0,1)$) for sparse recovery}\label{sec:lpl1}

For strong recovery, if $\ell_1$-minimization can recover all the
$k$-sparse vectors, then $\ell_p$-minimization is also guaranteed to
recover all the $k$-sparse vectors for all $p \in [0,1)$. However,
this does not necessarily indicate that the performance of
$\ell_p$-minimization ($0 \leq p<1$) is always better than that of
$\ell_1$-minimization. Example 1 in Section \ref{sec:example}
indicates that sometimes $\ell_1$-minimization can successfully
recover the original sparse vector while $\ell_p$-minimization ($p
\in (0,1)$) would return a vector that is denser than the original
vector. Moreover, our results for weak recovery indicates that the
performance of $\ell_1$-minimization is better than that of
$\ell_p$-minimization for all $p \in[0,1)$ in at least the large
$\alpha$ region ($\alpha
>0.9$).

\begin{figure}[ht]
\centering
\begin{tabular}{c c}
\includegraphics[scale=0.4]{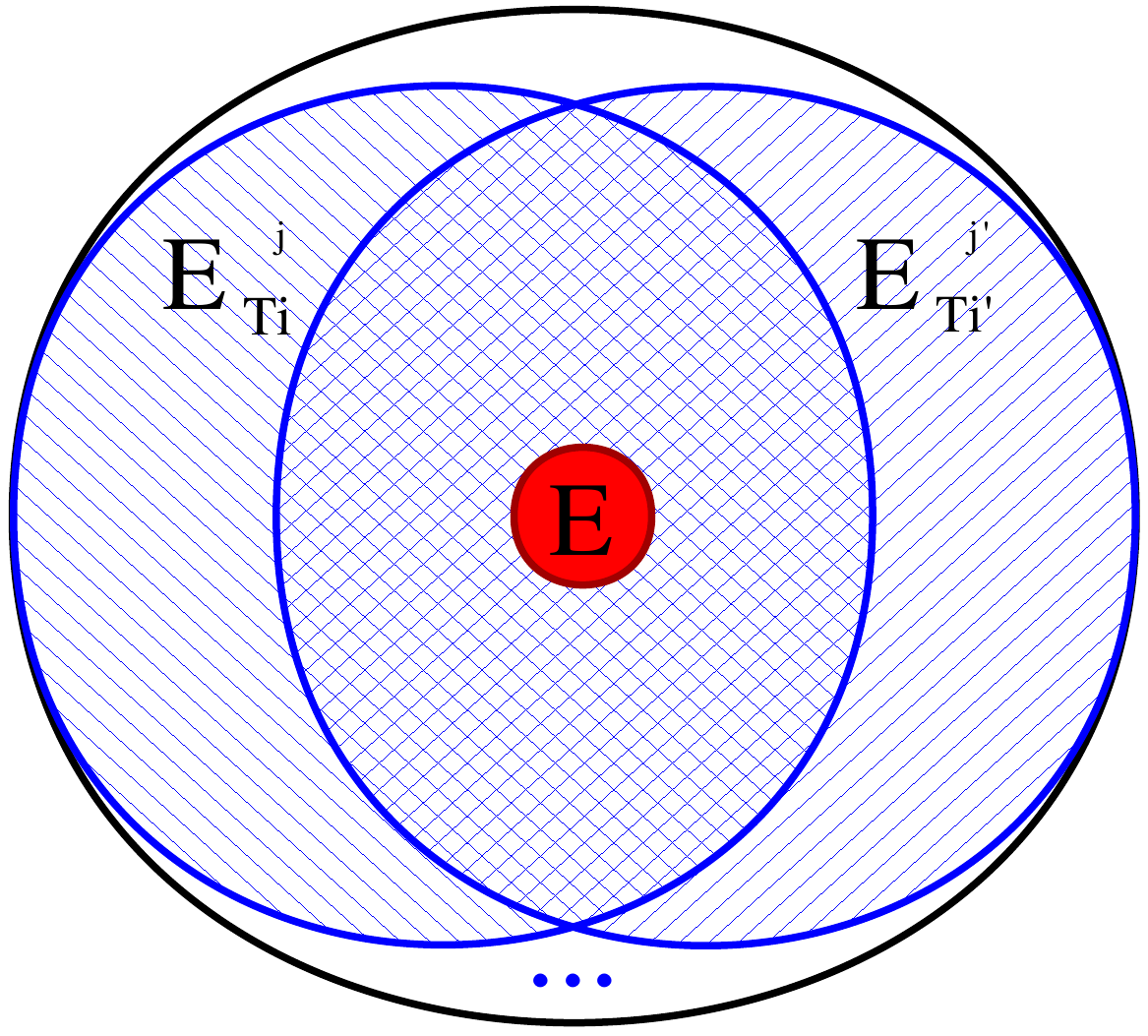}
&
\includegraphics[scale=0.4]{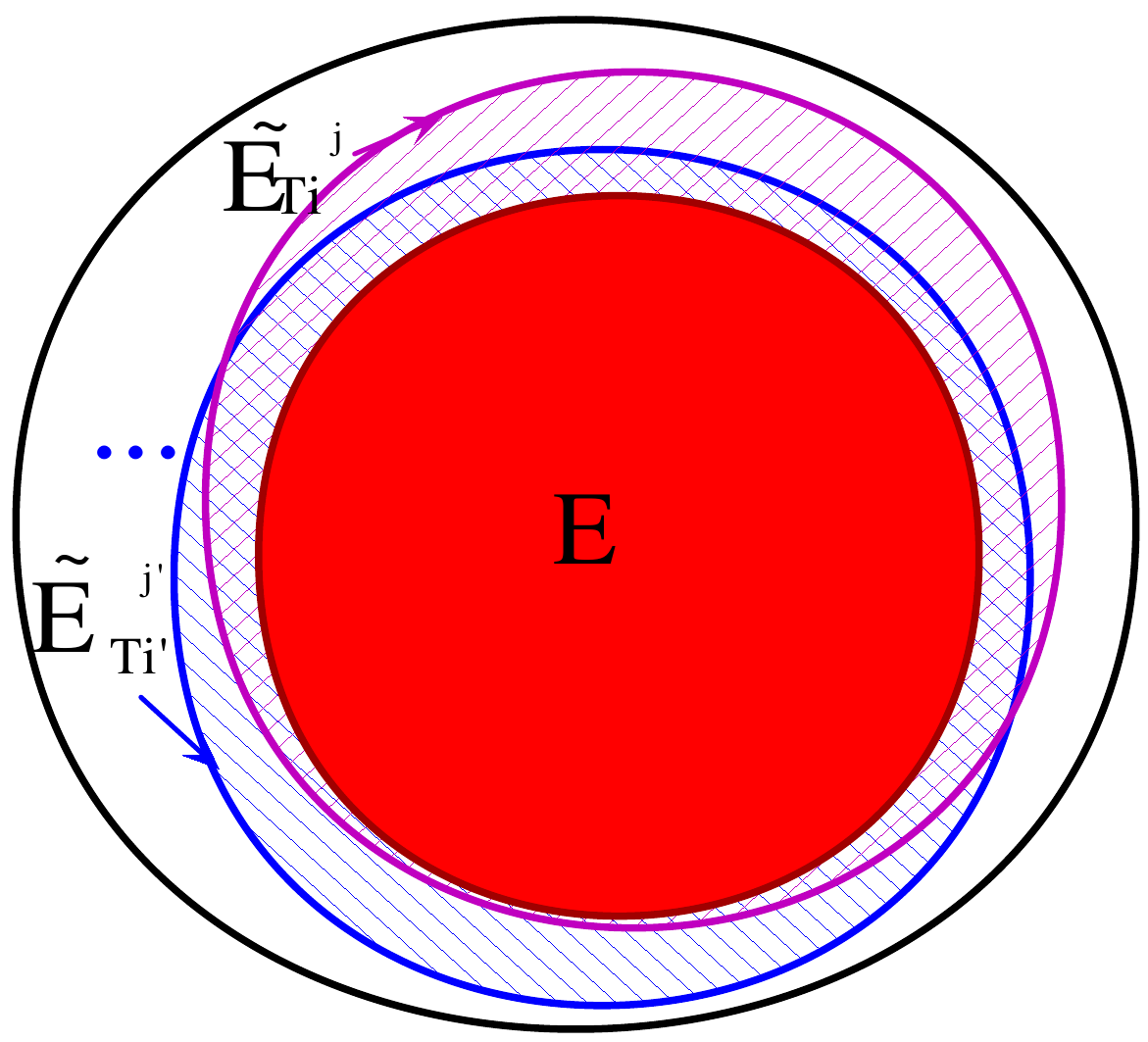}
\\
{\small (a) $\ell_1$-minimization }&  {\small (b) $\ell_p$-minimization}\\
\end{tabular}
      \caption{Comparison of $\ell_1$ and $\ell_p$-minimization for $ \rho \in (\rho_1, \rho_2)$.}
      \label{fig:rho1}
   \end{figure}

\begin{figure}[ht]
\centering
\begin{tabular}{c c}
\includegraphics[scale=0.4]{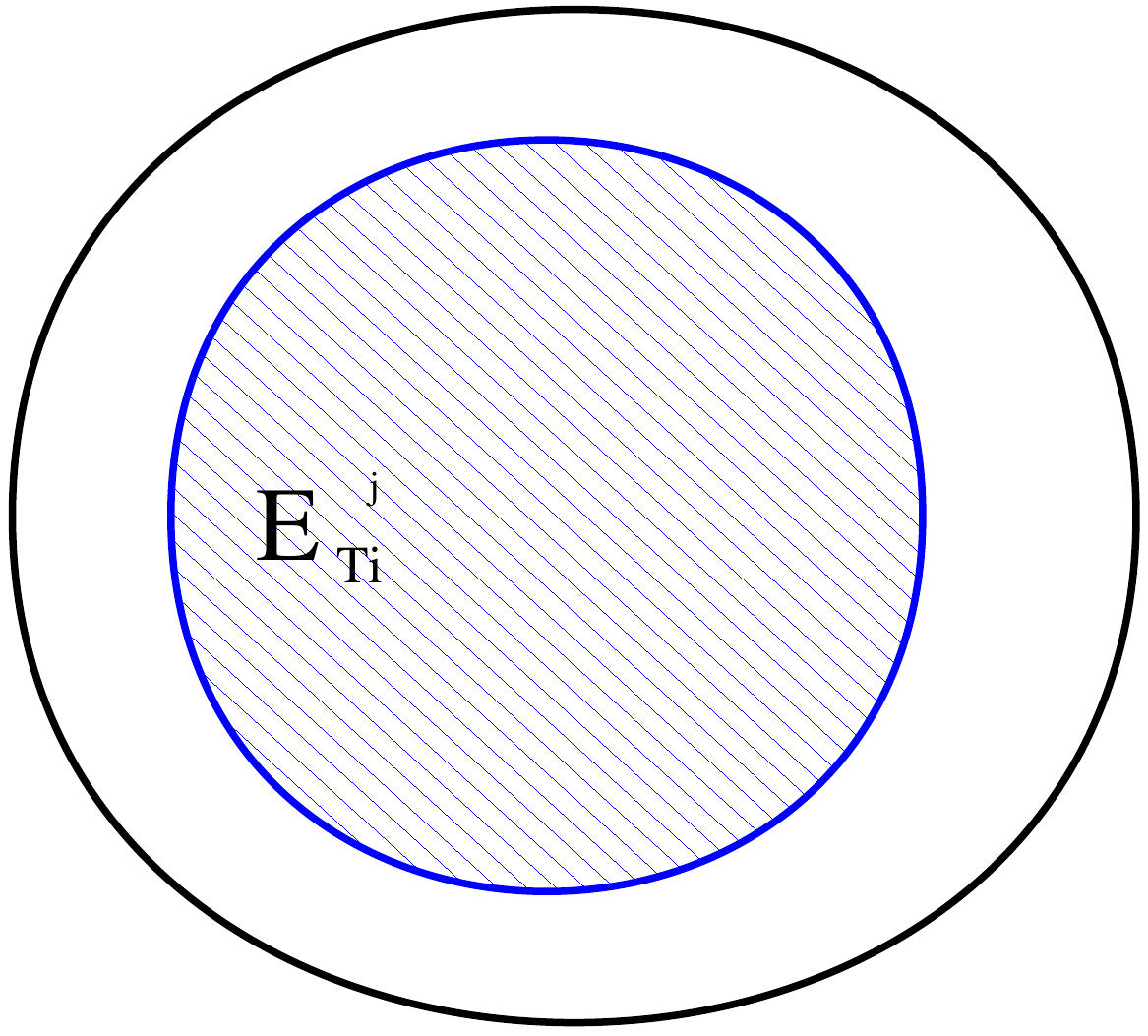}
&
\includegraphics[scale=0.4]{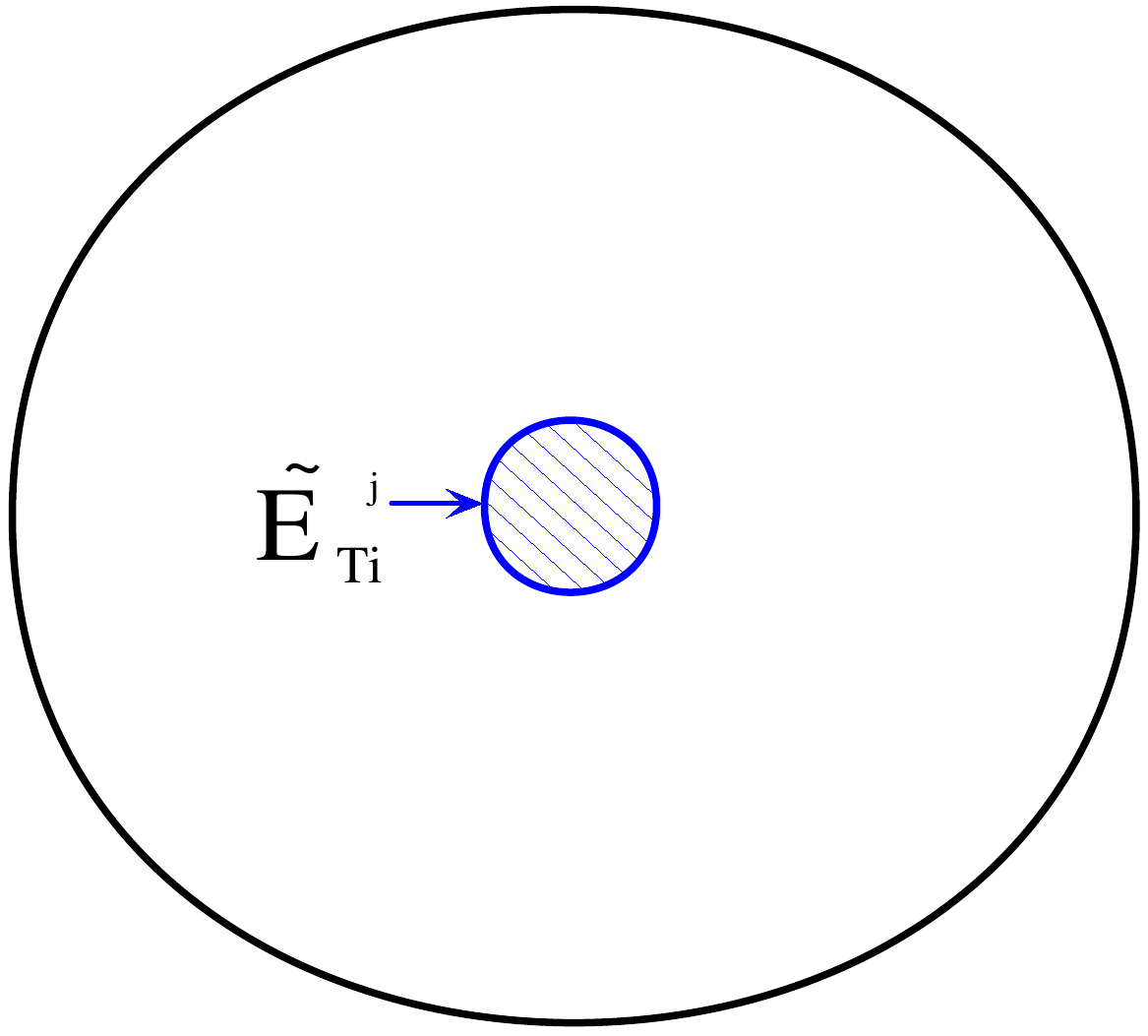}
\\
{\small (a) $\ell_1$-minimization }&  {\small (b) $\ell_p$-minimization}\\
\end{tabular}
      \caption{Comparison of $\ell_1$ and $\ell_p$-minimization for $ \rho \in (\rho_3, \rho_4)$.}
      \label{fig:rho2}
   \end{figure}

We can roughly interpret the result as follows. Let $\alpha<1$ be
very close to 1, let $n$ be large enough and $A$ is a random
Gaussian matrix. Then with overwhelming probability
$\ell_1$-minimization can recover all the vectors up to $\rho_1
n$-sparse and $\ell_p$-minimization with some $p \in [0,1)$ can
recover all the vectors up to $\rho_2 n$-sparse, and we know
 $\rho_1<\rho_2$ from our discussion on strong bound. Note that since the limiting
 threshold of strong recovery via $\ell_p$-minimization increases to 0.5 as $p$ goes to 0,
 then we have $\rho_1<\rho_2\leq 0.5$.
 However, if we only consider the ability to recover
all the vectors on one support with one sign pattern, with
overwhelming probability $\ell_1$-minimization can recover vectors
up to $\rho_3 n$-sparse, while $\ell_p$-minimization can recover
vectors up to $\rho_4 n$-sparse. From previous discussion about weak
recovery threshold, we know that when $\alpha$ is very close to 1,
$\rho_3>\frac{2}{3}>\rho_4 >\frac{1}{2}$. 
Therefore we have $\rho_3>\rho_4>\rho_2>\rho_1$. 
We illustrate the difference of $\ell_1$ and $\ell_p$-minimization
in Fig. \ref{fig:rho1} and Fig. \ref{fig:rho2}. 
Let $\Omega$ be the set of all $m \times n$ matrices with entries
drawn from standard Gaussian distribution, and the probability
measure $P(\Omega)=1$. We pick $\rho \in (\rho_1, \rho_2)$ in Fig.
\ref{fig:rho1}. For a random measurement matrix $A$ in $\Omega$,
since $\rho <\rho_3$, for any fixed support $T$ with $|T| =\rho n$
and any fixed sign pattern $\sigma_j$, with high probability
$\ell_1$-minimization can recover all the $\rho n$-sparse vectors on
$T_i$ with sign pattern $\sigma_j$. Since we also have
$\rho>\rho_1$, then with high probability strong recovery of
$\ell_1$-minimization fails, in other words, $\ell_1$-minimization
would fail to recover at least one vector with at most $\rho n$
non-zero entries. In Fig. \ref{fig:rho1} (a), $E_{T_i}^{\sigma_j}$
denotes the event that $\ell_1$-minimization can recover all the
$\rho n$-sparse vectors on support $T_i$ with sign patter
$\sigma_j$. Then $P(E_{T_i}^{\sigma_j})$ is very close to 1 for
every $i$ and $j$. There are ${n\choose \rho n}$ different supports,
and for each support, there are $2^{\rho n}$ different sign
patterns. Let $E$ denote the event that $\ell_1$-minimization can
recover all the $\rho n$-sparse vectors, then we have
\begin{equation*}
E=\bigcap \limits_{i \in \{1,...,{n\choose \rho n}\},  j \in
\{1,..., 2^{\rho n}\}} E_{T_i}^{\sigma_j}.
\end{equation*}
Then although $P(E_{T_i}^{\sigma_j})$ is the same for all $i$ and
$j$ and is very close to 1, $P(E)$ is close to 0, as indicated in
Fig. \ref{fig:rho1} (a). For $\ell_p$-minimization, since $\rho <
\rho_2$, then with high probability, $\ell_p$-minimization can
recover all the $\rho n$-sparse vectors. In Fig. \ref{fig:rho1} (b),
$\tilde{E}$ denotes the event that $\ell_p$-minimization can recover
all the $\rho n$-sparse vectors, then
\begin{equation*}
\tilde{E}=\bigcap \limits_{i \in \{1,...,{n\choose \rho n}\}, j \in
\{1,..., 2^{\rho n}\}} \tilde{E}_{T_i}^{\sigma_j},
\end{equation*}
where $\tilde{E}_{T_i}^{\sigma_j}$ denotes the event that
$\ell_p$-minimization recovers all the vectors on support $T_i$ with
sign pattern $\sigma_j$. In this case, $P(\tilde{E})$ is close to 1
as indicated in Fig. \ref{fig:rho1} (b). In Fig. \ref{fig:rho2}, we
pick $\rho \in (\rho_3, \rho_4)$. Then given any $i$ and $j$,
$\ell_1$-minimization can recover all the vectors on $T_i$ with sign
pattern $\sigma_j$ with high probability, while
$\ell_p$-minimization fails to recover at least one vector on $T_i$
with sign pattern $\sigma_j$ with high probability. Therefore
$P(E_{T_i}^{\sigma_j})$ is close to 1, while
$P(\tilde{E}_{T_i}^{\sigma_j})$ is close to 0 for any given $i$ and
$j$. Therefore, if the sparse vectors we would like to recover are
on one same support and share the same sign pattern,
$\ell_1$-minimization can be a better choice than
$\ell_p$-minimization for all $p \in [0,1)$ regardless of the
amplitudes of the entries of a vector.

To better understand how the recovery performance changes from
strong recovery to weak recovery, let us consider another type of
recovery: sectional recovery, which measures the ability of
recovering all the vectors on one support $T$. Therefore, the
requirement for successful sectional recovery is stricter than that
of weak recovery, but is looser than that of strong recovery. The
necessary and sufficient condition of successful sectional recovery
can be stated as:

\begin{theorem}\label{thm:sslp}
$\bfx$ is the unique solution to $\ell_p$-minimization problem ($p
\in [0,1]$) for all $\rho n$-sparse vector $\bfx$ on some support
$T$, if and only if
\begin{equation}\label{eqn:sslp}
\|B_{T}\bfz\|_p^p < \|B_{T^c}\bfz\|_p^p
\end{equation}
for all non-zero $\bfz \in \mathcal {R}^{n-m}$.
\end{theorem}

The difference of the null space condition for strong recovery and
sectional recovery is that (\ref{eqn:sslp}) should hold for every
support $T$ for strong recovery, but only needs to hold for one
specific support $T$ for sectional recovery. Though for strong
recovery, if the null space condition holds for $p \in [0,1]$, it
also holds for all $q \in [0,p]$, this argument is not true for
sectional recovery. Consider a simple example that the basis $B$ of
null space of $A$ contains only one vector in $\mathcal{R}^4$ and
$T=\{1,2\}$. If $B=[16, 16, 1, 36]$, then one can check that
$\|B_T\|_1=32<37=\|B_{T^c}\|_1$, but
$\|B_T\|^{0.5}_{0.5}=8>7=\|B_{T^c}\|_{0.5}^{0.5}$. If $B=[1, 4, 1,
9]$, then $\|B_T\|_1<\|B_{T^c}\|_1$, and
$\|B_T\|^{0.5}_{0.5}<\|B_{T^c}\|_{0.5}^{0.5}$. Therefore the null
space condition of successful sectional recovery holds for $p$ does
not necessarily imply that it holds for another $q \neq p$.

Following the technique in Section \ref{sec:wbd}, one can show that
when $\alpha \rightarrow 1$ and $n$ is large enough, the recovery
threshold of sectional recovery is 1/2 for all $p \in [0,1]$. We
skip the proof here as it follows the lines in Section
\ref{sec:wbd}. To summarize, regarding the recovery threshold when
$\alpha \rightarrow 1$, $\ell_p$-minimization ($p\in [0,1]$) has a
higher threshold for smaller $p$ for strong recovery; the threshold
is all 1/2 for all $p \in [0,1]$ for sectional recovery; and the
threshold is all 2/3 for $p \in [0,1)$ and 1 for $p=1$ for weak
recovery. We can see how recovery performance changes when the
requirement for successful recovery changes from strong to weak.

\section{Numerical Experiments}\label{sec:simu}
We present the results of numerical experiments to explore the
performance of $\ell_p$-minimization. As mentioned earlier,
(\ref{eqn:lp}) is indeed non-convex and it is hard to compute its
global minimum. Here we employ the iteratively reweighted least
squares algorithm \cite{IRLS}\cite{CY08} to compute the local
minimum of (\ref{eqn:lp}), please refer to \cite{CY08} about the
details of the algorithm.

\noindent \textbf{Example 2. $\ell_p$-minimization using IRLS
\cite{CY08}}

\begin{figure}
      \centering
      \includegraphics[scale=0.5]{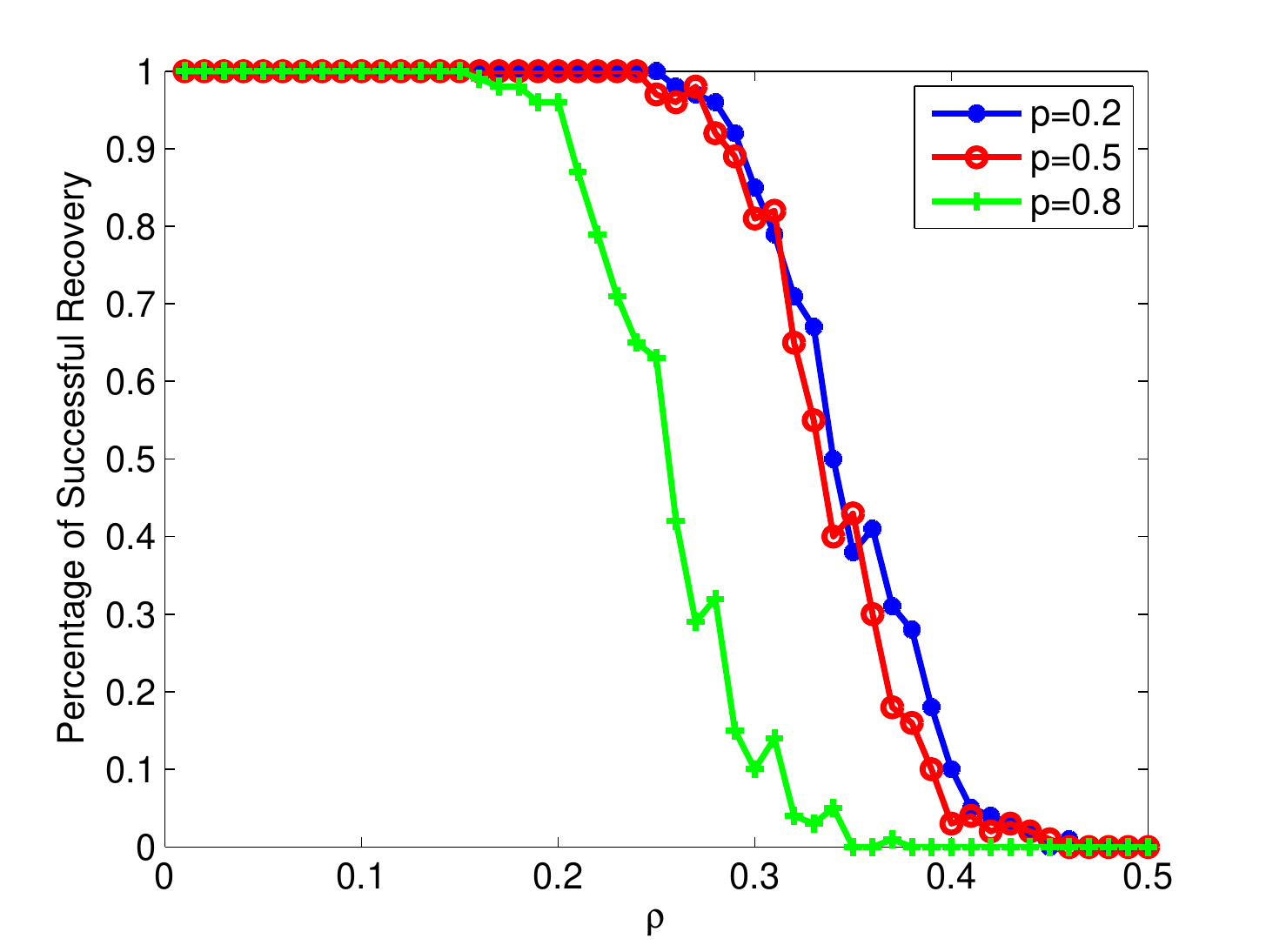}
 \caption{Successful recovery of $\rho n$-sparse vectors via $\ell_p$-minimization}
   \label{fig:simu}
\end{figure}

 We fix $n=200$ and $m=100$, and increase $\rho$ from 0.01 to
0.5 as a percentage of $n$. For each $\rho$, we repeat the following
procedure 100 times. We first generate a $n$-dimensional vector
$\bfx$ with $\rho n$ nonzero entries. The location of the non-zero
entries are chosen randomly, and each non-zero value follows from
standard Gaussian distribution. We then generate a $m \times n$
matrix $A$ with i.i.d. $\mathcal{N}(0,1)$ entries. We let
$\bfy=A\bfx$ and run the iteratively reweighted least squares
algorithm to search for a local minimum of (\ref{eqn:lp}) with $p$
chosen to be 0.2, 0.5, and 0.8 respectively. Let $\bfx^*$ be the
output of the algorithm, if $\|\bfx^*-\bfx\|_2 \leq 10^{-4}$, we say
the recovery of $\bfx$ is the successful. Figure \ref{fig:simu}
records the percentage of times that the recovery is successful for
different sparsity $\rho n$. Note that the iteratively reweighted
least squares algorithm is designed to obtain a local minimum of the
$\ell_p$-minimization problem (\ref{eqn:lp}), and is not guaranteed
to obtain the global minimum. However, as shown in Figure
\ref{fig:simu}, it indeed recovers the
sparse vectors up to certain sparsity.  
For $\ell_{0.2}$, $\ell_{0.5}$ and $\ell_{0.8}$-minimization
computed by the heuristic, the sparsity ratios of successful
recovery are 0.025, 0.024, and 0.015 respectively.

\begin{figure}
      \centering
      \includegraphics[scale=0.5]{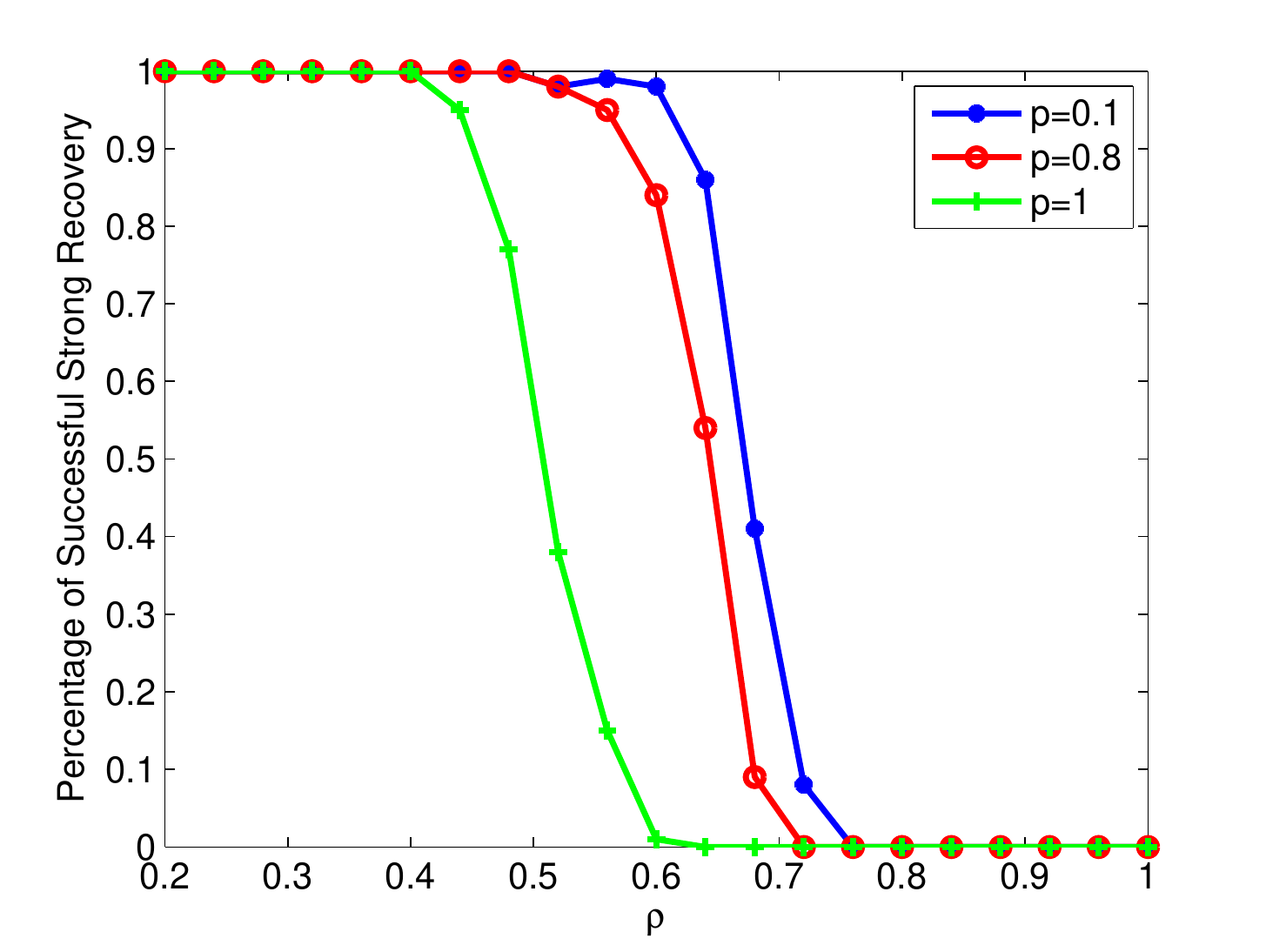}
 \caption{Successful strong recovery of $\rho n$-sparse vectors}
   \label{fig:strong}
\end{figure}

\begin{figure}
      \centering
      \includegraphics[scale=0.5]{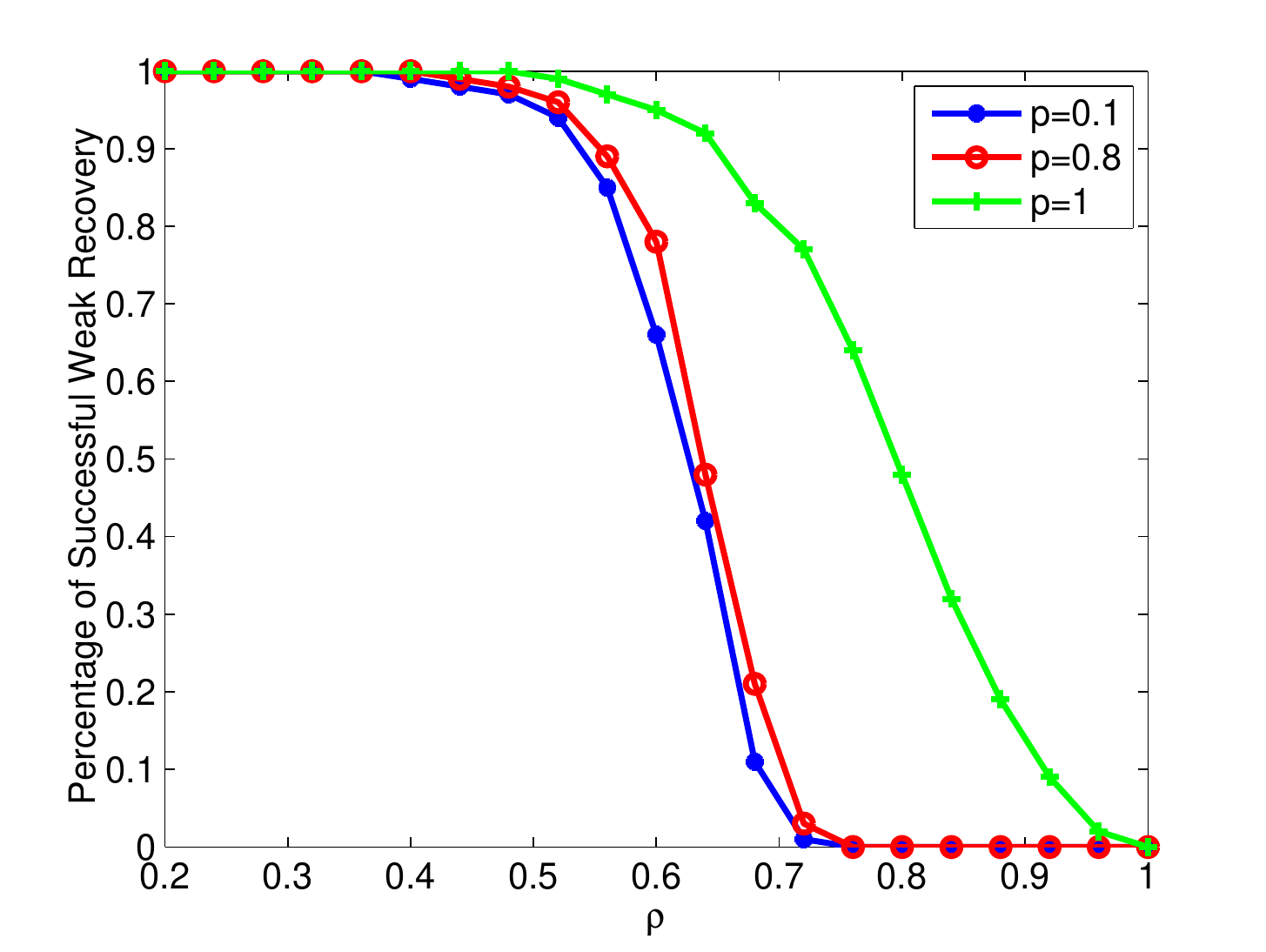}
 \caption{Successful weak recovery of $\rho n$-sparse vectors}
   \label{fig:weak}
\end{figure}

\noindent \textbf{Example 3. Strong recovery vs. weak recovery}

We also compare the performance of $\ell_{p}$-minimization and
$\ell_1$-minimization both for strong recovery in Fig.
\ref{fig:strong} and for weak recovery in Fig. \ref{fig:weak} when
$\alpha$ is large. We employ \texttt{CVX} \cite{CVX} to solve
$\ell_1$-minimization and still employ the iteratively reweighted
least squares algorithm to compute a local minimum of
$\ell_p$-minimization. We fix $n=50$ and $m=48$ and independently
generate one hundred random matrices $A^{m \times n}$ with i.i.d.
$\mathcal{N}(0,1)$ entries and evaluate the performance of strong
recovery and weak recovery. For each matrix, we increase $\rho$ from
0.04 to 1. In weak recovery, we consider recovering nonnegative
vectors on support $T=\{1,..., \rho n\}$. For a given $\rho$, we
generate one hundred and fifty vectors and claim the weak recovery
of $\rho n$-sparse vectors to be successful if and only if all the
vectors are successfully recovered. For each vector $\bfx$, $x_i$
($i \in T$) is generated from $\mathcal{N}(0,1)$ with probability
0.5, and $\mathcal{N}(1000,1)$ with probability 0.5. As discussed in
Section \ref{sec:null}, the condition for successful weak recovery
via $\ell_1$-minimization is the same for every nonnegative vector
on $T$, therefore if $\ell_1$-minimization recovers all the vectors
we generated, it should also recover all the nonnegative vectors on
$T$.
$\ell_p$-minimization ($p\in [0,1)$), on the other hand, can recover
some nonnegative vectors on $T$ while at the same time fails to
recover some other nonnegative vectors on $T$. 
Therefore, since we could not check every nonnegative $\bfx$ on $T$,
$\ell_p$-minimization ($p<1$) can still fail to recover some other
nonnegative vector on $T$ even if we declare the weak recovery to be
``successful''.
In strong recovery, for each $\rho$, 
we generate two hundred vectors and claim the strong recovery to be
successful if and only if all these vectors are correctly recovered.
To generate a $\rho n$-sparse vector $\bfx$, we first randomly pick
a support $T$ with $|T| =\rho n$. For each $x_i$ ($i \in T$), $x_i$
is generated from $\mathcal{N}(0,1)$ with probability 0.5, from
$\mathcal{N}(1000,1)$ with probability 0.25, and from
$\mathcal{N}(-1000,1)$ with probability 0.25.
The average performance of one hundred random matrices for strong
recovery is plotted in Fig. \ref{fig:strong}, and the average
performance of weak recovery is plotted in Fig. \ref{fig:weak}. Note
that we only apply iteratively reweighted least squares algorithm to
approximate the performance of $\ell_p$-minimization, therefore the
solution returned by the algorithm may not always be the solution of
$\ell_p$-minimization. Simulation results indicate that
 for
strong recovery, the recovery threshold increases as $p$ decreases,
while for the weak recovery, interestingly, the recovery threshold
of $\ell_1$-minimization is higher than any other
$\ell_p$-minimization for $p<1$.

\section{Conclusion}\label{sec:conclusion}

This paper analyzes the ability of $\ell_p$-minimization ($0\leq
p\leq 1$) to recover high-dimensional sparse vectors from
low-dimensional linear measurements where the measurement matrix
$A^{m \times n}$ has i.i.d. standard Gaussian entries. When $\alpha=
m/n \rightarrow 1$, we provide a tight threshold $\rho^*(p)$ of the
sparsity ratio separating the success and failure of strong recovery
which requires to recover all the sparse vectors. $\rho^*(p)$
strictly decreases from 0.5 to 0.239 as $p$ increases from 0 to 1.
For weak recovery which only needs to recover sparse vectors on some
support with some sign pattern, we first provide an equivalent null
space characterization of successful weak recovery, then prove that
the threshold of sparsity ratio separating the success and failure
of $\ell_p$-minimization is 2/3 for all $p<1$, compared with the
threshold 1 for $\ell_1$-minimization. For any $\alpha<1$, we
provide a bound $\rho^*(\alpha,p)$ of sparsity ratio below which
strong recovery via $\ell_p$-minimization succeeds with overwhelming
probability, and our bound $\rho^*(\alpha,p)$ improves on the
existing bounds in the large $\alpha$ region. We also provide a
bound $\rho^*_w(\alpha,p)$ of sparsity ratio below which weak
recovery succeeds with overwhelming probability.

Throughout the paper, we assume that the measurements $\bfy=A\bfx$
are exact, and it would be interesting to consider the case that the
measurements are noisy, i.e. $\bfy=A\bfx +\bfe$ where $\bfe$ is the
vector of noise. Moreover, we assume that $\bfx$ is exactly sparse,
i.e. most of its entries are exactly zero. The extension of results
to approximately sparse vectors whose coefficients (if ordered)
decay rapidly is also worth pursuit.

\vspace{0.09in}\noindent {\bf Acknowledgments:} 
The research is supported by NSF under CCF-0835706.

\bibliographystyle{IEEEtranS}

\appendix

\subsection{Calculation of  $\lambda_{\max}(\alpha,p)$ in Lemma
7}\label{sec:lemma7}
Define $c_{\max}=\frac{1}{n}\max_{ \bfz \in \mathcal{S}}
\|B\bfz\|_p^p$, then
for any non-zero vector $\bfz$, $\|B\bfz\|_p^p \leq \|\bfz\|_p^p
c_{\max}n$. Let $\Sigma_1$ be a $\gamma$-net of $\mathcal{S}$ with
cardinality at most $(1+2/\gamma)^{n-m}$ \cite{Ledoux01} and
$\gamma>0$ to be chosen later, and define
\begin{equation}\nonumber
\eta = \frac{1}{n}\max_{ \bfz \in \Sigma_1} \|B\bfz\|_p^p.
\end{equation}
Then from the definition of $\gamma$-net, for every $\bfz \in
\mathcal{S}$, there exists $\bfz' \in \Sigma_1$ such that
$\|\bfz-\bfz'\|_2 \leq \gamma$. Note that for every $\bfz \in
\mathcal{S}$,
$\|B\bfz\|_p^p \leq  \|B\bfz'\|_p^p +\|B(\bfz-\bfz')\|_p^p  \leq
\eta n + \gamma^p c_{\max} n$. Then $c_{\max}n \leq \eta n +
\gamma^p c_{\max} n$, which leads to
%
\begin{equation}\label{eqn:netmax}
c_{\max} \leq \eta/(1-\gamma^p).
\end{equation}

To characterize $c_{\max}$, we first characterize $\eta$. We will
show that there exists a constant $a>E[|X|^p]$ where $X \sim
\mathcal{N}(0,1)$ such that with overwhelming probability,
$\|B\bfz\|_p^p < an$ for all $\bfz$ in $\Sigma_1$. Given $\bfz \in \mathcal{S}$, 
$B_i\bfz$ ($i=1,...,n$) are i.i.d. $\mathcal{N}(0,1)$ random
variables where $B_i$ is the $i^{\textrm{th}}$ row of $B$. Then
\begin{eqnarray}
&&P(\eta \geq a) =P(\exists \bfz \in \Sigma_1 \textrm {s.t. }
\|B\bfz\|_p^p \geq an) \nonumber\\
& \leq & \sum_{ \bfz \in \Sigma_1 }P(\|B \bfz\|_p^p \geq an) \nonumber\\
& \leq & (1+2/\gamma)^{n-m} \min_{t>0} e^{-tan} E[e^{t\sum_i
|B_i\bfz|^p}]\nonumber\\
& =& (1+2/\gamma)^{(1-\alpha)n}\min_{t>0} e^{-tan}E[e^{t|X|^p}]^n\nonumber\\
&=&
e^{\large((1-\alpha)\log(1+\frac{2}{\gamma})+\min_{t>0}(\log(E[e^{t|X|^p}])-at)\large)n},
\label{eqn:eta}
\end{eqnarray}
where $X \sim \mathcal{N}(0,1)$, the first inequality follows from
the union bound, and the second inequality follows from the Chernoff
bound.

Since the second-order derivative of $\log(E[e^{t|X|^p}])-at$ to $t$
is positive, then its minimum is achieved where its first-order
derivative is 0. To calculate the value of $t$ where the minimum is
achieved, we have
\begin{eqnarray}
0&=&\frac{d[\log(E[e^{t|X|^p}])-at]}{dt} \nonumber \\
&=&\frac{d}{dt}(\log(\sqrt{\frac{2}{\pi}}\int_0^{\infty}e^{tx^p-\frac{1}{2}x^2}dx)-at) \nonumber \\
&=& \frac{\int_0^{\infty}x^p e^{tx^p-\frac{1}{2}x^2} dx}{
\int_0^{\infty}e^{tx^p-\frac{1}{2}x^2 }dx}-a \label{eqn:minimum}.
\end{eqnarray}
Note that when $a>E[|X|^p]$, the solution of $t$ to
(\ref{eqn:minimum}) is always positive,  thus it is also the
solution to $\min_{t>0}(\log(E[e^{t|X|^p}])-at)$. One can check that
 for any $\gamma$, the exponent in (\ref{eqn:eta}) is negative when
$a$ is large enough. To see this, let
$t=2(1-\alpha)\log(1+2/\gamma)/a$, then $\log(E[e^{t|X|^p}])-at$
goes to $-2(1-\alpha)\log(1+2/\gamma)$ as $a$ goes to infinity.
Thus, when $a$ is sufficiently large,
$\log(E[e^{t|X|^p}])-at<-(1-\alpha)\log(1+2/\gamma)$ if $t=c/a$.
Therefore, the exponent in (\ref{eqn:eta}) is negative when $a$ is
large enough. Thus, we can pick $a(\alpha,p,\gamma)$ large enough
such that there exists some constant $c_{12}>0$ and
$P(\eta \geq a(\alpha,p,\gamma)) \leq e^{-c_{12}n}$ holds.
Then 
\begin{equation} \nonumber
P( c_{\max}\geq \frac{a(\alpha,p,\gamma)}{1-\gamma^p} )\leq
P(\frac{\eta}{1-\gamma^p}\geq
\frac{a(\alpha,p,\gamma)}{1-\gamma^p})\leq e^{-c_{12}n},
\end{equation}
where the first inquality follows from (\ref{eqn:netmax}). 
Let
\begin{equation}\label{eqn:lambdamax}
 \lambda_{\max}(\alpha,p)=\min_{\gamma}a(\alpha,p,\gamma)/(1-\gamma^p),
\end{equation}
then there exists $c_{12}(\alpha, p, \lambda_{\max})>0$ such that
with probability at least $1-e^{-c_{12}n}$, for every $\bfz \in
\mathcal{S}$, $\|B\bfz\|_{p}^{p} < \lambda_{\max}n$. Thus, Lemma 7
follows.


\subsection{Proof of Lemma 8}

\begin{proof}
Define
$c'_{\max}= \frac{1}{(1-\rho) n}\max_{ \bfz \in \mathcal{S}}
\|B_{T^c}\bfz\|_p^p.$ 
Let $\Sigma_4$ be a $\gamma$-net of $\mathcal{S}$ with cardinality
at most $(1+2/\gamma)^{n-m}$ and $\gamma$ being the value where
$\lambda_{\max}(\frac{\alpha-\rho}{1-\rho},p)$ is achieved, and
define
\begin{equation}\nonumber
\eta' = \frac{1}{(1-\rho)n}\max_{ \bfz \in \Sigma_4} \|B\bfz\|_p^p.
\end{equation}
Then same as that in the calculation of $\lambda_{\max}(\alpha ,p)$
in Appendix-\ref{sec:lemma7}, we have
\begin{equation}\nonumber
c'_{\max} \leq \eta'/(1-\gamma^p).
\end{equation}
We use $\lambda_{\max}$ to denote
$\lambda_{\max}(\frac{\alpha-\rho}{1-\rho},p)$ for simplicity.
 We first show that 
with overwhelming probability, $\|B_{T^c}\bfz\|_p^p <
(1-\rho)\lambda_{\max}n$ for all $\bfz$ in $\mathcal{S}$, or equivalently $c'_{\max} < \lambda_{\max}$. Note that 
\begin{eqnarray}
&&P( c'_{\max}\geq \lambda_{\max} ) \nonumber\\
&\leq&
P(\eta'/(1-\gamma^p)\geq \lambda_{\max}) \nonumber\\
&=&P(\exists \bfz \in \Sigma_4 \textrm { s.t. }
\|B_{T^c}\bfz\|_p^p \geq (1-\rho)\lambda_{\max}(1-\gamma^p)n) \nonumber\\
& \leq
& \sum_{ \bfz \in \Sigma_4} P(\|B_{T^c}\bfz\|_p^p \geq (1-\rho)\lambda_{\max}(1-\gamma^p)n) \nonumber \\
& \leq & (1+\frac{2}{\gamma})^{n-m} \min_{t>0} \frac{
E[e^{t\sum_{i\in T_c}
|B_i\bfz|^p}]}{e^{t(1-\rho)\lambda_{\max}(1-\gamma^p)n}}\nonumber\\
& =& (1+\frac{2}{\gamma})^{(1-\alpha)n}\min \limits_{t>0} \frac{E[e^{t|X|^p}]^{(1-\rho)n}}{e^{t(1-\rho)\lambda_{\max}(1-\gamma^p)n}}\nonumber\\
&=&
e^{(1-\rho)n\Large(\frac{1-\alpha}{1-\rho}\log(1+\frac{2}{\gamma})+\min
\limits_{t>0}(\log(E[e^{t|X|^p}])-\lambda_{\max}(1-\gamma^p)t)\Large)},
\label{eqn:eta'}
\end{eqnarray}
where $X \sim \mathcal{N}(0,1)$. 
From the definition of
$\lambda_{\max}(\frac{\alpha-\rho}{1-\rho},p)$, and that $\gamma$ is
chosen to be the value where
$\lambda_{\max}(\frac{\alpha-\rho}{1-\rho},p)$ is achieved, we know
that there exists $c_{13}>0$ such that (\ref{eqn:eta'}) $\leq
e^{-c_{13}n}$. Therefore it holds with probability at least
$1-e^{-c_{13}n}$ that for all $\bfz \in \mathcal{S}$,
$\|B_{T^c}\bfz\|_{p}^{p} < (1-\rho)\lambda_{\max}n$.

Similarly, define
$c'_{\min}= \frac{1}{(1-\rho)n}\min_{ \bfz \in \mathcal{S}}
\|B\bfz\|_p^p$. 
Let $\Sigma_5$ be a $\gamma$-net of $\mathcal{S}$ with cardinality
at most $(1+2/\gamma)^{n-m}$ and $\gamma$ being the value where
$\lambda_{\min}(\frac{\alpha-\rho}{1-\rho},p)$ is achieved, note
that
\begin{equation}\nonumber
\lambda_{\min}(\frac{\alpha-\rho}{1-\rho},p)=\gamma^{p(\frac{1-\alpha}{1-\rho}+\epsilon)}-\gamma^p\lambda_{\max}(\frac{\alpha-\rho}{1-\rho},p)
\end{equation}
for some $\epsilon \in (0, \frac{1-\alpha}{1-\rho})$ according to
the definition of $\lambda_{\min}(\frac{\alpha-\rho}{1-\rho},p)$. We
use $\lambda_{\min}$ and $\lambda_{\max}$ to denote
$\lambda_{\min}(\frac{\alpha-\rho}{1-\rho},p)$ and
$\lambda_{\max}(\frac{\alpha-\rho}{1-\rho},p)$ for simplicity. We
define
\begin{equation}\nonumber
\theta'= \frac{1}{(1-\rho)n}\min_{ \bfz \in \Sigma_5}
\|B_{T^c}\bfz\|_p^p.
\end{equation}
Like in the calculation of $\lambda_{\min}(\alpha, p)$ in Section
\ref{sec:lambda}, we have
\begin{equation} \nonumber\label{eqn:netmin'}
c'_{\min} \geq \theta'-\gamma^pc'_{\max}.
\end{equation}

We next show that 
with overwhelming probability, $\|B_{T^c}\bfz\|_p^p >
(1-\rho)\lambda_{\min}n$ for all $\bfz$ in $\mathcal{S}$, or
equivalently $c'_{\min} > \lambda_{\min}$. Note that
\begin{eqnarray}
&&P(c'_{\min}\leq \lambda_{\min}) \nonumber\\
&=&P( c'_{\min}\leq
\gamma^{p(\frac{1-\alpha}{1-\rho}+\epsilon)}-\gamma^p \lambda_{\max}) \nonumber\\
&\leq&  P( \theta'-\gamma^pc'_{\max}\leq
\gamma^{p(\frac{1-\alpha}{1-\rho}+\epsilon)}-\gamma^p
\lambda_{\max})
\nonumber\\
&\leq& P(\theta' \leq \gamma^{p(\frac{1-\alpha}{1-\rho}+\epsilon)})+P(c'_{\max} \geq \lambda_{\max}) \nonumber\\
& \leq & P(\theta' \leq
\gamma^{p(\frac{1-\alpha}{1-\rho}+\epsilon)})+ e^{-c_{13}n},
\label{eqn:c'min}
\end{eqnarray}
where the last inequality follows from (\ref{eqn:eta'}). To
calculate $P(\theta' \leq
\gamma^{p(\frac{1-\alpha}{1-\rho}+\epsilon)})$,
note that
\begin{eqnarray}
&&P(\theta' \leq \gamma^{p(\frac{1-\alpha}{1-\rho}+\epsilon)}) \nonumber\\
&=& P(\exists \bfz \in \Sigma_5 \textrm{ s.t. }
\|B_{T^c}\bfz\|_p^p \leq (1-\rho)\gamma^{p(\frac{1-\alpha}{1-\rho}+\epsilon)}n) \nonumber\\
& \leq & \sum_{ \bfz \in \Sigma_5}P(\sum_{i \in T^c} |B_i \bfz|^p \leq (1-\rho)\gamma^{p(\frac{1-\alpha}{1-\rho}+\epsilon)}n) \nonumber\\
& \leq & (1+\frac{2}{\gamma})^{(1-\alpha)n} e^{(1-\rho)n}E[e^{-\gamma^{-p(\frac{1-\alpha}{1-\rho}+\epsilon)}|X|^p}]^{(1-\rho)n} \nonumber\\
&=&
e^{(1-\rho)n\Large(\frac{1-\alpha}{1-\rho}\log(1+\frac{2}{\gamma})+\log(E[e^{-\gamma^{-p(\frac{1-\alpha}{1-\rho}+\epsilon)}|X|^p}])+1\Large)}
\nonumber\\
&=&e^{(1-\rho)n\Large(\frac{1-\alpha}{1-\rho}\log(1+\frac{2}{\gamma})+\log(O(\gamma^{\frac{1-\alpha}{1-\rho}+\epsilon}))+1\Large)},
 \label{eqn:theta'}
\end{eqnarray}
where $X \sim \mathcal{N}(0,1)$, the second inequality follows from
the Chernoff bound, and the last equality follows from
(\ref{eqn:bigo}). Since $\gamma$ is chosen to be the value where
$\lambda_{\min}(\frac{\alpha-\rho}{1-\rho},p)$ is achieved, then
according to the definition of
$\lambda_{\min}(\frac{\alpha-\rho}{1-\rho},p)$, (\ref{eqn:theta'})
$\leq e^{-\kappa n}$ for some positive $\kappa>0$. Thus, from
(\ref{eqn:c'min}) we have
\begin{equation}\nonumber
P(c'_{\min}\leq \lambda_{\min}) \leq e^{-\kappa n}+ e^{-c_{13}n}
\leq e^{-c_{14}n},
\end{equation}
for some $c_{14}>0$. Then, with probability at least
$1-e^{-c_{14}n}$, for all $\bfz \in \mathcal{S}$,
$\|B_{T^c}\bfz\|_p^p
>(1-\rho)\lambda_{\min}(\frac{\alpha-\rho}{1-\rho},p)n$.
\end{proof}

\subsection{Calculation of $\tilde{\lambda}_{\max}(\alpha,p , \rho)$ in Lemma 9}

\begin{proof}
Define
$\tilde{c}_{\max}= \frac{1}{\rho n}\max_{ \bfz \in \mathcal{S}}
\|B_{T^-}\bfz\|_p^p$. 
Let $\Sigma_6$ be a $\gamma$-net of $\mathcal{S}$ with cardinality
at most $(1+2/\gamma)^{n-m}$ and $\gamma>0$ to be chosen later, and
define
$
\tilde{\eta} = \frac{1}{\rho n}\max_{ \bfz \in \Sigma_4}
\|B_{T^-}\bfz\|_p^p.
$
Then from (\ref{eqn:expansion}), for any $\bfz \in \mathcal{S}$,
$\bfz=\sum_{j\geq 0} \gamma_j \bfv_j$ hold, where $\gamma_0=1$,
$\gamma_j \leq \gamma^j$ and $\bfv_j \in \Sigma_6$. From
(\ref{eqn:BT-}) we have
\begin{eqnarray}
 \|B_{T^-}\bfz\|_p^p&\leq & \sum \limits_{j\geq 0}  \gamma^{jp} \sum \limits_{i \in T:
(B_i\bfv_j)x_i < 0}  |B_i\bfv_j|^p \nonumber\\
& \leq &  \sum \limits_{j\geq 0}  \gamma^{jp} \tilde{\eta}\rho n \nonumber \\
& \leq & \tilde{\eta}\rho n/(1-\gamma^p) \label{eqn:BT-up}
\end{eqnarray}

Since (\ref{eqn:BT-up}) holds for every $\bfz \in \mathcal{S}$, then
$\tilde{c}_{\max}\rho n \leq \tilde{\eta}\rho n/(1-\gamma^p)$,
which leads to
%
$\tilde{c}_{\max} \leq \tilde{\eta}/(1-\gamma^p)$. 
Define a random variable $S_i$ for each $i$ in $T$ that is equal to
1 if $B_i \bfz  x_i<0$ and equal to 0 otherwise. Then
$\|B_{T^-}\bfz\|_p^p =\sum_{i\in T} |B_i\bfz|^pS_i$. 
Then for any $\tilde{a}$,
\begin{eqnarray}
&&P( \tilde{c}_{\max}\geq \frac{\tilde{a}}{1-\gamma^p} )\leq
P(\frac{\tilde{\eta}}{1-\gamma^p}\geq \frac{\tilde{a}}{1-\gamma^p})
\nonumber\\ &=&P(\tilde{\eta} \geq \tilde{a}) =P(\exists \bfz \in
\Sigma_6 \textrm { s.t. }
\|B_{T^-}\bfz\|_p^p \geq \tilde{a}\rho n) \nonumber\\
& \leq & \sum_{ \bfz \in \Sigma_6 }P(\|B_{T^-} \bfz\|_p^p \geq \tilde{a}\rho n) \nonumber\\
& = & (1+\frac{2}{\gamma})^{n-m}  P(\sum_{i \in T} |B_i\bfz|^pS_i \geq \tilde{a}\rho n) \nonumber\\
& \leq & (1+\frac{2}{\gamma})^{(1-\alpha)n} \min_{t>0} \frac{ E[e^{t|X|^pS}]^{\rho n}}{e^{t\tilde{a}\rho n}}\nonumber\\
&=&
e^{\large((1-\alpha)\log(1+\frac{2}{\gamma})+\rho\min_{t>0}(\log(E[e^{t|X|^pS}])-\tilde{a}t)\large)n},
\label{eqn:etat}
\end{eqnarray}
where $X \sim \mathcal{N}(0,1)$, $S=1$ if $X <0$ and $S=0$
otherwise.

Since the second derivative of $\log(E[e^{t|X|^pS}])-\tilde{a}t$ to
$t$ is positive, then its minimum is achieved where its first
derivative is 0. To calculate the value of $t$ where the minimum is
achieved, we have
\begin{eqnarray}
0&=&\frac{d[\log(E[e^{t|X|^pS}])-\tilde{a}t]}{dt} \nonumber \\
&=&\frac{d}{dt}(\log(\sqrt{\frac{1}{2\pi}}\int_0^{\infty}e^{tx^p-\frac{1}{2}x^2}dx+\frac{1}{2})-\tilde{a}t) \nonumber \\
&=& \frac{\int_0^{\infty}x^p e^{tx^p-\frac{1}{2}x^2} dx}{
\int_0^{\infty}e^{tx^p-\frac{1}{2}x^2 }dx+\sqrt{\pi/2}}-\tilde{a}
\label{eqn:minimum'}.
\end{eqnarray}
Note that when $\tilde{a} > E[|X|^pS]$, the solution of $t$ to
(\ref{eqn:minimum'}) is always positive, thus it is also the
solution to $\min_{t>0}(\log(E[e^{t|X|^pS}])-\tilde{a}t)$. Given any
$\rho$ and $\gamma$, when $\tilde{a}$ is large enough, the exponent
in (\ref{eqn:etat}) is negative. We can pick $\tilde{a}(\alpha, p,
\rho,\gamma)$
as small as possible while still keeping the exponent in (\ref{eqn:etat}) negative. 
Let
\begin{equation}\label{eqn:tildelambdamax}
\tilde{\lambda}_{\max}(\alpha, p, \rho)=\min_{\gamma}
\frac{\tilde{a}(\alpha, p, \rho,\gamma)}{1-\gamma^p},
\end{equation}
then there exists $c_{15}>0$ such that with probability at least
$1-e^{-c_{15}n}$, $c_{\max}<\tilde{\lambda}_{\max}(\alpha, p,
\rho)$, or equivalently, for every $\bfz \in \mathcal{S}$,
$\|B_{T^-}\bfz\|_{p}^{p} < (1-\rho)\tilde{\lambda}_{\max}(\alpha, p,
\rho)n$. Thus, Lemma 9 follows.

\end{proof}

\end{document}